\documentclass[usenatbib,letter]{mn2e}
\usepackage{epsfig}

\def\lsim{\lower0.6ex\vbox{\hbox{$ \buildrel{\textstyle <}\over{\sim}\ $}}}
\def\gsim{\lower0.6ex\vbox{\hbox{$ \buildrel{\textstyle >}\over{\sim}\ $}}}

\def\lb{\left[}
\def\rb{\right]}
\def\lp{\left(}
\def\rp{\right)}
\def\lb{\left[}
\def\rb{\right]}
\def\beq{\begin{equation}}
\def\eeq{\end{equation}}
\def\beqa{\begin{eqnarray}}
\def\eeqa{\end{eqnarray}}

\def\Msun{M_{\odot}}

\def\kms{{\rm km} \ {\rm s}^{-1}}
\def\cm3{{\rm cm^{-3}}}
\def\cm2{{\rm cm^{-2}}}
\def\gyr{\rm Gyr}

\def\etarho{\eta_{\rm d}}
\def\etap{\eta_P}
\def\fb{f_b}
\def\tcool{\tau_{\rm c}}
\def\rcool{R_{\rm c}}
\def\ccool{C_{\rm c}}
\def\rhocool{\rho_{\rm c}}
\def\ncool{n_{\rm c}}

\def\rs{R_{\rm s}}
\def\rvir{R_{\rm v}}
\def\cvir{C_{\rm v}}
\def\vvir{V_{\rm v}}
\def\vmax{V_{\rm max}}
\def\rhovir{\bar{\rho}_{\rm v}}
\def\Mvir{M_{\rm v}}

\def\masscool{M_{\rm c}}
\def\masscl{M_{\rm cl}}
\def\massgal{M_{\rm g}}
\def\masshot{M_{\rm h}}
\def\massrcool{M_{\rcool}}

\def\mcloud{m_{\rm cl}}
\def\rcl{r_{\rm cl}}
\def\vcl{\rm v_{cl}}

\def\rhow{{\rho}_w}
\def\rhowarm{\rho_w}
\def\ncloud{\phi_{\rm cl}}
\def\tcl{\theta_{\rm cl}}

\def\L23{\Lambda_{23}}
\def\Lzz{\Lambda_{z}}
\def\Lz{\Lzz}
\def\m6{m_6}
\def\mue{\mu_e}
\def\mup{\mu_i}
\def\Tw{T_{\rm w}}
\def\Th{T}

\def\gal {\lower .0ex\hbox{\rlap{\raise .0ex\hbox{\hskip .0ex
        {\ifmmode{\scriptstyle \bullet}\else
                {$\scriptstyle \bullet$}\fi}}}
        \hskip -.75ex{\ifmmode{\scriptstyle \sim}\else
                {$\scriptstyle \sim$}\fi}}}

\def\gal{\tiny{_{200}}}

\def\t9{t_{_{8}}}
\def\T6{T_{_{6}}}
\def\Lt{(\Lz\t9)}

\title{Multi-Phase Galaxy Formation: High Velocity Clouds
and the Missing Baryon Problem}

\author[A. H. Maller and J. S. Bullock]
{Ariyeh H. Maller$^1$, James S. Bullock$^{2,3,*}$\\
$^1$Astronomy Department, University of Massachussets at Amherst,
LGRT-B 619E, 710 North Pleasant St., Amherst, 01003; \\ ari@astro.umass.edu
\\
$^2$Harvard Smithsonian Center for Astrophysics, MS-51,Cambridge, MA 02138; jbullock@cfa.harvard.edu \\
$^3$Department of Physics and Astronomy, 4129 FRH, University of
California, Irvine, CA 92697-4574; bullockj@uci.edu\\
$^*$Hubble Fellow
}

\begin{document}

\maketitle

\begin{abstract}
The standard treatment of cooling  in  Cold Dark Matter halos  assumes
that all  of the gas within a  ``cooling radius''  cools and contracts
monolithically to fuel galaxy  formation.   Here we take into  account
the   expectation  that the  hot gas  in  galactic halos  is thermally
unstable and prone  to fragmentation during cooling  and show that the
implications are more far-reaching  than previously expected: allowing
multi-phase cooling fundamentally alters expectations about gas infall
in  galactic  halos  and naturally  gives  rise  to   a characteristic
upper-limit on the masses of galaxies,  as observed.  Specifically, we
argue that cooling should  proceed via the formation of  high-density,
$\sim 10^4$K clouds, pressure-confined within   a hot gas  background.
The background medium that emerges has a  low density, and can survive
as a hydrostatically stable  corona  with a long  cooling   time.  The
fraction of halo baryons contained in  the residual hot core component
grows with halo  mass because the cooling  density  increases with gas
temperature,  and this leads    to an upper-mass limit in   quiescent,
non-merged galaxies of $\sim 10^{11} \Msun$.  

In this scenario,
galaxy formation is fueled by the infall of pressure-supported clouds.
For Milky-Way-size systems,  clouds of  mass $\sim 5\times10^6  \Msun$
that formed or merged within the last  several Gyrs should still exist
as a residual population in  the halo, with a total  mass in clouds of
$\sim 2 \times  10^{10} \Msun$.  The   baryonic mass of the Milky  Way
galaxy is explained naturally in  this model, and is  a factor of  two
smaller than would result  in the standard treatment without feedback.
We  expect clouds in  galactic halos to be  $\sim 1$kpc in size and to
extend $\sim 150$kpc  from galactic centers.  The predicted properties
of   Milky  Way clouds   match    well the  observed radial   velocity
distribution, angular sizes, column densities,  and velocity widths of
High  Velocity Clouds  around our  Galaxy.  The clouds  we predict are
also  of the  type needed  to explain  high-ion  absorption systems at
$z<1$, and the predicted  covering factor around external galaxies  is
consistent with observations.
\end{abstract}


\section{Introduction}
Cooling and galaxy formation within  dark matter  halos was
first\footnote{Their ideas were based on those of
\citet{binn:77,ro:77} and \citet{silk:77} and were applied to CDM
specifically by \citet{bfpr:84}} discussed  in a modern 
context by \citet{wr:78}, who
argued that gas cooling was a main driver behind
the characteristic mass of galaxies. After halo
collapse, gas is assumed to shock-heat to the halo temperature, 
and to cool over a characteristic timescale
\beq
\label{eq:cool_time}
\tcool \simeq {{k_b T}\over{n_i \Lambda(T)}}, 
\eeq
which depends on the particle number density of the ionized gas $n_i$, 
the gas temperature $T$,
the Boltzman constant $k_b$, 
and the cooling function $\Lambda(T)$.
\citet{wf:91} extended this approach in order to make predictions as a function
of time and  position  in a halo.   The framework assumes
that all of the  gas
within  a central, high-density ``cooling radius''  cools and falls in
to fuel  galaxy assembly, while gas beyond this   radius remains hot.  This
cooling-radius method for tracking gas cooling is certainly a 
useful approach, and it  has  become the  basis for gas  accretion
estimates  in   semi-analytic  galaxy  formation  models  
\citep[recently,][]{spf:01,bens:03,hatt:03,hs:03,ny:04}.    Hydro-dynamical
simulations seem to verify this picture, at least roughly
\citep{katz:92,tw:95,swtk:01,yssw:02,hell:03b}.

Implicit  in this  standard  model is that  all  of the  gas within the
cooling radius cools and contracts monolithically. 
It has been known for some time, 
however, that the  hot gas associated with galaxies
(and clusters) should be thermally unstable and prone to fragmentation
\citep{field:65,fr:85,ml:90}. At a given radius within a halo, 
the cooling time can 
increase for  some gas  and decrease  for other  gas as density
and temperature differences become enhanced by the cooling process.
The result is
a fragmented distribution of cooled material, in
the form  of warm ($\sim 10^4$K)   clouds, pressure-supported within a
hot gas background.  In this paper we argue that the consequences of including
this expected ingredient  may be far-reaching.  The  residual hot
gas   core  has   a    low  density, and   thus      can exist  as   a
pressure-supported  corona for a  long  time without cooling.   The
fraction of  baryonic mass contained in  the hot core  component grows
with halo velocity (or temperature) and we show  below that this gives
rise to a characteristic cooled, central  galaxy mass of $\sim 10^{11}
\Msun$ in  high-mass  halos.  Interestingly, this  is  roughly what is
needed   to explain the bright-end   cutoff  in the galaxy  luminosity
function.  Similarly,  including this  multi-phase treatment can  help
explain the    masses  of Milky-Way type galaxies   without  the need for
excessive feedback.

In our picture, the gas supply into galaxies is governed by
the infall of warm clouds.  We
suggest  that the  cloud population  will have a velocity
dispersion similar to that of the host halo, and that clouds will fall
in to feed galaxy  formation only when  cloud-cloud collisions or  ram
pressure forces rob  them of angular  momentum and  kinetic energy.  In
Milky-Way-size halos, the residual population of clouds is expected to
be substantial.  The clouds occupy typical galacto-centric radii of
$\sim 100$kpc, and can explain
the High Velocity Cloud
(HVC) population around the  Milky Way and high-ion  absorption
systems in external   galaxies.  A  characteristic cloud
mass  of $\sim 5 \times 10^6  \Msun$ matches
most of the observed  properties of HVCs
(see  \S   \ref{sec:hvc}).  The  same
characteristic  cloud   mass  is consistent    with our    theoretical
expectations 
(\S  \ref{sec:cloud-masses}), aides  in the understanding
of absorption systems (\S \ref{sec:abs}),  and helps explain the total
mass of the Milky Way without any significant feedback
(\S\ref{sec:infall}).  Based on this evidence, we argue
that there is direct observational support for the idea that
gas fragmentation be included in models of CDM-based galaxy formation.

In  what follows we  will use the Milky Way  galaxy halo as a fiducial
case for  comparison.  The properties of  our  ``Milky Way''  dark
matter halo  are adopted from results  of \citet{kzs:02}.  The
authors use a
wide variety  of Galactic  data and  a baryonic-infall  calculation to
determine a best-fit halo mass of $\Mvir  \simeq 10^{12} \Msun$ and an
initial halo  maximum circular velocity of  $\vmax = 163 \kms$.  Their
Milky Way mass is  motivated by the  mass models of \citet{db:98}, who
obtain $M_{G} = (4-6) \times 10^{10} \Msun$.  

The results of \citet{kzs:02} and \citet{db:98} also
provide a  useful illustration of the
``over-cooling   problem'' faced by  the
standard treatment of cooling in galaxy halos. 
In the standard model, the
baryons that end up in the Galaxy are
simply those that exist within the cooling radius:
$\masscool = f_c \fb \Mvir$.  Here $f_c \simeq 0.7$ is the fraction of
baryonic mass within the cooling radius  for Milky-Way type halos (see
\S \ref{sec:hot_phase})  and $\fb = \Omega_b/\Omega_m \simeq 0.17$  is the cosmic baryon
fraction \citep{sper:03}.  Based on the
numbers quoted above, the ratio  of the ``expected''  cooled mass
to the actual  mass of the  Galaxy, $f_{\rm G} = M_{\rm G}/\masscool$,
is significantly less than unity:
\beqa
\label{eq:galfrac}
f_{\rm G}  \simeq 0.43 \lb \frac{M_{\rm G}}{5 \times 10^{10} \Msun}\rb
\lb \frac{\Mvir}{10^{12} \Msun}\rb^{-1}
\lb \frac{\fb f_c}{0.12 }\rb^{-1}.
\eeqa
If the standard cooling arguments are correct,  then less than half of
the baryons that have cooled onto the Galaxy  still exist there today.
If feedback is to explain this, it  requires that the Galaxy lost half
of its mass via strong winds without destroying the thin disk.  In our
picture, this  difficult series of events  is avoided because  a large
fraction  of the mass  within the  cooling  radius never fell  in, but
remains in  the halo   in the  form  of  a   warm/hot medium  (see  \S
\ref{sec:infall}).  This effect  becomes  more important  in high-mass
halos because the density  of the hot  gas core (which scales like the
cooling density) increases with halo temperature.

We  have made an effort  to frame our results as   an extension of the
standard treatment   of   cooling, and  we  give  comparisons   to the
single-phase approach whenever  possible.  Of course, gas cooling  and
accretion in galactic  halos is more  complicated than  the 
static
halo   model  we use.   Indeed gas falling  into  halos may   or  may  not be
shock-heated efficiently,  and some  gas  was likely accreted  as cold
material, either stripped   from infalling  satellites, or simply   as
``cold flows'' \citep{bd:03,keres:04}.   Our goal is  merely to extend
the  standard semi-analytic treatment   to   include an allowance    for
multi-phase gas during cooling and to work out the implications
of this approach.
  We expect that  no matter how gas clouds are
accreted,   our  qualitative  expectations     will  hold.   Detailed,
high-resolution
hydrodynamic  simulations will be
needed to test these expectations.  Unfortunately, as we discuss in \S
\ref{sec:conc}, the numerical challenges facing   such an endeavor may  be
significant.

\begin{table*}
\begin{center}
\begin{tabular}{ccp{12cm}}
\hline
Symbol & Equation where first used & Description\\
\hline
\hline
\hline
$ \fb$ & (\ref{eq:galfrac})
& The fraction of mass in
 the universe in the form of baryons. \\
\hline
$\Mvir, \rvir, \vvir$ & (\ref{eq:vir})(\ref{eq:vir})(\ref{eq:vir})
& Halo virial properties: mass, radius, and  velocity.\\
\hline
$\rs, \vmax, \cvir$ &
(\ref{eq:nfw}) (\ref{eq:sis}) (\ref{eq:mass_growth})
& Halo parameters: NFW
scale radius, maximum circular velocity, and
NFW concentration.\\
\hline
$\Lambda(T), \Lz $  & (\ref{eq:rhocool}) (\ref{eq:Lfit})
& The cooling function and its approximate metallicity scaling\\
\hline
$\rho_c, \ncool, \rcool $  & (\ref{eq:rhocool}) (\ref{eq:ncool})
(\ref{eq:rcooldef})
& The cooling density, corresponding electron number density, and cooling
radius\\
\hline
$\mup, \mue, Z_g$ &
(\ref{eq:temp}) (\ref{eq:rhocool})
 (\ref{eq:Lfit})
& The mean gas mass per particle,
mean gas mass per electron, and the gas metallicity.\\
\hline
$\eta_T, \eta_d, \eta_P$&  (\ref{eq:eta}) (\ref{eq:eta}) (\ref{eq:eta})
&Ratios of average hot gas core temperature, density, and pressure to
the values at the cooling radius. \\
\hline
$\rhow, \Tw$
& (\ref{eq:rhow}) (\ref{eq:rhow})
&  Warm cloud density and temperature.\\
\hline
$\mcloud,\rcl,\vcl$
& (\ref{eq:rhow}) (\ref{eq:rcl}) (\ref{eq:ram})
&  Cloud mass, cloud radius, and the typical cloud velocity.\\
\hline
$\masscool,\masshot, \masscl, \massgal$ & (\ref{eq:mcool})
(\ref{eq:mcool}) (\ref{eq:taucc}) (\ref{eq:massbudget})
& The total mass in various phases: cooled material, hot core, clouds,
central galaxy.\\
\hline
$\tau_{\rm ram},\tau_{cc}, \tau_{\rm in}$ &  (\ref{eq:tauram})
(\ref{eq:taucc})      (\ref{eq:dmdt})
& Cloud population time scales: the ram-pressure time, the cloud-cloud
collision time, the cloud infall time.\\
\hline
\end{tabular}
\caption{Frequently used symbols
}\label{tab:symbols}
\end{center}
\end{table*}

Before providing an outline of the paper, we mention
that in a series of papers, D. Lin and collaborators 
\citep{ml:90,ml:92,lm:92,murr:93,bl:00,lm:00,ml:04} have explored the fate of 
warm clouds in a hot gas medium.  The analysis that follows
builds on their
work.  \citet{mm:96} consider the presence of pressure-supported
clouds in dark matter halos in order to explain
QSO absorption line systems.  We make a similar connection in \S\ref{sec:abs}.

The next section contains a review of the properties of dark matter halos.
In \S\ref{sec:hot_phase} we describe the standard treatment of 
radiative cooling in halos.  \S\ref{sec:two-phase} extends 
this model to include the formation of warm clouds within a background 
hot gas medium. In \S\ref{sec:cloud-masses} we explore cloud masses,
bracket the allowed range, and discuss mass scales of interest.
\S\ref{sec:infall} is devoted to modeling galaxy fueling via cloud infall.
In \S\ref{sec:hvc} and \S\ref{sec:abs} we compare our expected cloud
populations to HVC data and CIV absorption system observations, respectively.
\S\ref{sec:lf} presents an explanation for the exponential cutoff in the 
bright-end of the galaxy luminosity function.
Future directions and implications are discussed in 
\S\ref{sec:imp} and we summarize in \S\ref{sec:conc}.  
In what follows, we adopt a flat $\Lambda$CDM 
cosmology, with parameters set by the best-fit WMAP values:
$h=0.72$, $\Omega_b h^2=0.024$, and $\Omega_m h^2 = 0.14$
\citep[][see also \cite{prim:02}]{sper:03}
The implied fraction of mass in baryons is $f_b = 0.17$.

\section{Dark Matter Halos}
\label{sec:dm}

A dark matter halo of a given mass $\Mvir$
is characterized by a  virial radius
$\rvir$.  The spherical top hat model \citep{gg:72} provides a
reasonable estimate of the value of $\rvir$, which is
set by the radius at which the average mass enclosed 
equals a characteristic virial density 
$\rhovir \equiv \Delta_{\rm v} \rho_{u}$.
Here $\rho_{u}$ is the matter density of the universe and 
$\Delta_{\rm v}$ is a cosmology-dependent variable that
can vary as a function of redshift.  For our adopted
$\Lambda$CDM cosmology  
$\Delta_{\rm v} \simeq 360/(1+z)$ when $z \lsim 1$
and $\Delta_{\rm v} \simeq 178$ when $z \gsim 1$ \citep[a more precise
fit is given by][]{bn:98}.
With this definition the halo virial mass and radius are related
via
\footnote{The virial radius and velocity scale with mass in the following way\\
$\rvir \simeq 206 \rm{h}^{-1}\rm{kpc} 
\lp \frac{\Delta_{\rm v} \Omega_m}{97.2}\rp^{-1/3} 
\lp \frac{\Mvir}{10^{12} h^{-1} M_{\odot}}\rp^{1/3} (1+z)^{-1}$\\
$\vvir \simeq 144 {\rm km \,\, s} ^{-1}
\lp \frac{\Delta_{\rm v} \Omega_m}{97.2}\rp^{1/6}
\lp \frac{\Mvir}{10^{12} h^{-1} M_{\odot}}\rp^{1/3} (1+z)^{1/2}$
}
\beq
\label{eq:vir}
\rvir= \lp{{3\Mvir}\over{4\pi \rhovir}}\rp^{1/3}, \quad 
\vvir = \sqrt{\frac{G \Mvir}{\rvir}}.
\eeq

The singular isothermal sphere (SIS) is a simple approximation that is
often adopted for the 
density profile of a dark matter halo:
\beq
\label{eq:sis}
\rho(R)={{\rhovir \rvir^2}\over{3R^2}}
={{\vmax^2}\over{4\pi G }}{{1}\over{R^2}}.
\eeq
The SIS profile has a rotation curve that is flat as a function of $R$:
$V(R) = \vvir = \vmax$, where $\vmax$ is defined to be
the maximum rotation velocity of the halo.
The temperature of a singular isothermal sphere is related to its
velocity by
\beq
\label{eq:temp}
T={{\mup m_p c_g^2}\over{\gamma k_b}} =10^6 K \lp{{\vmax}\over{163 \kms}}\rp^2,
\eeq
where $c_g = \vmax/\sqrt{2}$ is the sound speed of the gas, 
$m_p$ is the proton mass,
the polytropic index is $\gamma=1$ for an isothermal gas,
and $\mup = 0.62$ is the mean mass per particle (electrons and nucleons) of 
the ionized gas in units of the proton mass assuming a $30\%$ mass fraction 
in Helium.

While the SIS is convenient for illustrative purposes, 
a better fit to the results of cosmological N-body simulations
is the NFW profile \citep{nfw:95,kkbp:01}
\beq
\label{eq:nfw}
\rho(R)={{\rho_s \rs^3}\over{R(R+\rs)^2}},
\eeq
where $\rho_s$ is a characteristic density.
The scale radius, $\rs$ is often expressed in terms of the
concentration $\cvir=\rvir/\rs$.  Given a halo mass, the value of
$\cvir$ sets the value of $\rho_s$, and the profile
is determined.  Simulations show that the median value
of $\cvir$ for a halo of mass $\Mvir$ at redshift $z$
is well-approximated by
$\cvir(M,z) \simeq 9.6 (\Mvir/M_*)^{-0.13} (1+z)^{-1}$ 
\citep{bull:01a}.  Here $M_*$ $\simeq 10^{13} \Msun$ 
is the characteristic mass for collapse at $z=0$ 
for our cosmology \citep[see][]{lc:93}.

The  maximum circular velocity for an NFW profile occurs at a radius
$R_{\rm max} \simeq 2.15 \rs$, where 
$\vmax^2 \equiv G M(R_{\rm max})/R_{\rm max}$.  For
our adopted cosmology, a 
good fit to the virial velocity in terms of $\vmax$ is
\beq
\vvir \simeq 0.468 \vmax^{1.1},
\eeq
which is good to $1\%$ for $\vvir$ between $80 \kms$ and 
$1200 \kms$.
We assume that in the absence of cooling,
the relationship between velocity and temperature
for NFW halos follows that for isothermal halos (equation \ref{eq:temp}).

Finally, N-body simulations show   that 
the mass accretion history of a dark matter halo as a function
of redshift $z$ follows a remarkably 
well-defined function of the final concentration $\cvir^0$
\citep{wech:02}:
\beq
\label{eq:mass_growth}
\Mvir(z) = \Mvir(0) \exp\lb\frac{-8.2z}{\cvir^0}\rb.
\eeq
Galaxy-mass halos have $\cvir^0 \simeq 13$, 
and typically have accreted half of their mass by
$z_f \sim 1.1$, corresponding to a lookback ``formation time''
of $t_f \simeq 8 \gyr$.

As discussed in \citet{wech:02} dark halos  tend  to  grow  qualitatively
from the inside out \citep[see also][]{hws:03,zb:03,tkgk:03}.  The
central  density remains roughly constant at   the value set during an
early, rapid accretion phase of halo buildup.  Because  of this, for a
given halo, $\vmax$ is relatively constant, as  a function of lookback
time.  Indeed for halos with $\vmax \simeq 175-250 \kms$ at $z=0$
simulations show that $\vmax$ stays approximately
constant in the main progenitor back to  $z \simeq z_f$ (R. Wechsler,
private communication).  Motivated by these   results, in what
follows we will  assume that the halo $\vmax$ (and thus its temperature)
will remain constant back to the  time of formation.
With these properties of dark matter halos defined we can move on 
to the treatment of gas in a halo.

\section{Hot Gas and Collisional Cooling}
\label{sec:hot_phase}

In the standard picture of CDM-based galaxy formation, gas collapses 
with the dark matter, and subsequently shock heats to the 
temperature of  the virialized halo \citep[see][]{wf:91}.  
The result is an extended halo of hot gas, which begins
to cool over a characteristic timescale (equation \ref{eq:cool_time}), 
and provides the gas reservoir for galaxy formation.  
Recently it has been shown in hydrodynamical simulations that the
situation is not as simple as this \citep{bd:03,keres:04}.  In 
low-mass halos, $\vmax \lsim 100 \kms$, the cooling time of the gas 
is so short that a shock can not be maintained.  Thus gas accretes
in a "cold flow".  For halos with $\vmax \gsim 100 \kms$ simulations 
find that the gas does indeed shock heat to the temperature of the 
halo.  In our multi-phase model of the gas, we can interpret 
these cold flows as gas that simply enters the halo as warm clouds 
(without first shock-heating and subsequently re-forming clouds).

In what follows we work within the framework of the standard
picture and investigate how the treatment of fragmentary cloud
cooling will change the results.  We do so mainly in the
spirit of direct comparison.  Although, as we show below,
most of the interesting ramifications of including 
multi-phase cooling occur in high-mass halos, where
shock-heating is expected to occur, and the standard picture
is roughly valid.
As  mentioned in the introduction,  different
expectations  for  the  nature of   gas accretion onto  halos  and the
efficiency of shock heating will affect  our results somewhat, but the
qualitative nature of our conclusions will not change.
For example, gas stripped from infalling halos will likely
fragment into clouds during this process, as  seen directly in the
the Magellanic Stream \citep[e.g.][]{ww:96} and lead to a
configuration similar to what we discuss below (although via
a different chain of events).  Of course, more direct
modeling will be needed to test these expectations in detail.

\begin{figure} %fig1
\centering 
\vspace{0pt} 
\epsfig{file=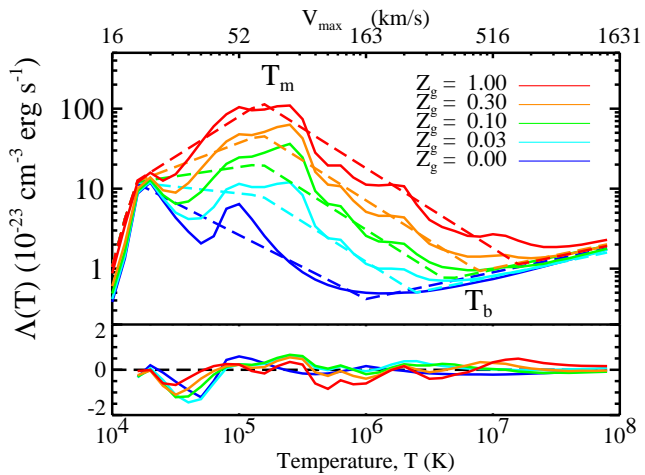,width=\linewidth} 
\vspace{0pt}
\caption{
The  cooling function  
for five different metallicities as a function of halo
temperature (lower scale) and corresponding halo
maximum velocity (upper scale). The five metallicities, from top
to bottom, are: $Z_g = 1.0, 0.3, 0.1, 0.03$ and $0.0$ relative to
solar.  Also shown are 
simple power-law fitting functions described in Appendix \ref{sec:coolfunc} 
for each metallicity (dashed lines). The power laws change slope at
two characteristic temperatures, $T_m$ and $T_b$, corresponding to when
metal line and Bremsstrahlung cooling start to dominate, respectively.  
The fractional 
difference between the full function and the fit are shown in the bottom 
panel.
}\label{fig:coolfunc}
\end{figure}

\subsection{The Initial Hot Gas Profile}

After halo formation,  we assume that  the hot gas obtains an 
extended density   profile.  
Motivated by the non-radiative  hydrodynamic simulations summarized in
\citet{frenk:99}, we assume that a dark halo with an NFW
profile and  concentration
$\cvir$ will initially have hot gas that
traces the DM at   large  radius, but that develops a  thermal core at  
$\simeq 3\rs/4$:
\beq
\label{eq:init}
\rho_g^i(R) = \frac{\rs^3 \rho_0}{(R + \frac{3}{4}\rs)(R + \rs)^2},
\quad \rho_0 = {{M_b}\over{4 \pi \rs^3 g(\cvir)}}.
\eeq
This profile gives a  good fit to Fig. 12 in  
\citet{frenk:99}.  The 
normalization,  $\rho_0$,  is set so that   the total initial gas mass
within the  halo is $M_b = f_b  \Mvir$.  The function $g(x)$ describes
the radial gas mass profile
\beq
M_g^i(R) = M_b \frac{g(R/\rs)}{g(\cvir)},
\eeq
where
\beq
g(x)\equiv 9\ln{(1+ \frac{4}{3}x)} -8\ln(1+x)-\frac{4x}{1+x}.
\eeq
Note that the \citet{cole:00} group adopt a similar starting
point for their hot gas profile (a  non-singular  isothermal
configuration).  However, it has been common in other semi-analytic
models  to assume that the hot
gas  profile  mirrors  that  of  the dissipationless  dark matter halo
profile, $\rho_g^i(R) = f_b \rho_{dm}(R)$.  
We find that significant 
differences between this approach (assuming an NFW profile) and
the thermal-core assumption 
arise only for halos of cluster mass and above (where the
cores become large).  

\subsection{The Cooling Density}
Once the initial hot gas profile is in place,
the time it takes for gas to cool is only dependent
on its density and the rate of cooling, parameterized by
the cooling function $\Lambda(T)$.
Since cooling is triggered 
by collisional excitation, higher density material generally
cools first.
Given a time since the halo formed, $t_f$, the ``cooling density''
is the characteristic density above which gas can cool:
\beq
\label{eq:rhocool}
\rhocool = {{3 \mue^2 m_p k_b T}\over{2 \mup t_f \Lambda(T,Z_g)}}.
\eeq
Assuming a 30\% 
mass fraction in Helium\footnote{
A gas with $30\%$  Helium by mass has 3 Helium atoms for
every 28 Hydrogen atoms.  If the gas is 
fully ionized, there are roughly 34 electrons for every 
$31$ nuclei, with a mean mass per particle of $\mup \simeq 8/13 = 0.62$ 
and a mean mass per electron of $\mue \simeq 20/17 = 1.18$.} 
, the mean mass per particle in ionized gas is $\mup = 0.62$ and the mean
mass per electron is $\mue = 1.18$.
In what follows we will associate the time $t_f$ with the halo
formation time, which can be estimated using equation \ref{eq:mass_growth}.

The cooling function, $\Lambda(T)$, can be
calculated  as a function of  gas 
temperature, $T$, and metallicity $Z_g$ \citep{sd:93}.  We plot $\Lambda(T)$ 
as a function of temperature 
for several different gas metallicities as solid lines
in Figure \ref{fig:coolfunc}.
The top axis shows the halo velocity that 
corresponds to temperature value shown on the bottom axis.
The dashed lines show a series of power-law
fitting functions presented in Appendix \ref{sec:coolfunc}.

Usefully, as long as the gas is only mildly enriched, $Z_g \gsim 0.1$, 
then the cooling function in galaxy-size halos (with $T_m < T< T_b$)
is well-described by a simple power-law form
\beq
\label{eq:Lfit}
\Lambda(T,Z_g) \simeq 2.6 \times 10^{-23}  \quad \Lzz 
\lb{{T}\over{10^6 K}}\rb^{-1} \quad{\rm cm}^{3} {\rm erg} \,{\rm s}^{-1} .   
\eeq
The parameter $\Lzz$ is a constant that varies with the metallicity
of the gas. 
The lower limit on the range of validity corresponds to the temperature
where metal line cooling starts to dominate, $T_m \simeq 1.5 \times 10^5$,
and the upper limit 
is set by the temperature when Bremsstrahlung becomes the dominate
cooling process $T_{b} \simeq 10^6$K $+1.5 \times  {Z_{g}}^{2/3} 10^7$K.
 We will typically concern ourselves with
mildly enriched gas with  $Z_g = 0.1$, in which case $\Lzz = 1.0$,
and the above
expression is valid for halos with maximum velocities in the range 
$\sim 60 - 300 \kms$.
We comment  that our metallicity choice is consistent with metallicity 
estimates for some HVCs \citep{tripp:03,semb:04}.
For other gas metallicities the values of $\Lzz$ 
and the ranges of validity can be found in  Table \ref{tab:cf}.

It will be useful to express the cooling density in terms of
the corresponding electron number density
 $\ncool=\rhocool/(\mu_e m_{\rm p})$. Adopting equation (\ref{eq:Lfit})
for $\Lambda(T)$, we can derive a typical value
for $\ncool$ using equation~\ref{eq:rhocool}:
\beq
\label{eq:ncool}
\ncool \simeq 6.1 \times 10^{-5} {\rm cm}^{-3} \quad \T6^2 \Lt^{-1}.
\eeq
We have scaled our results by the characteristic temperature and formation time
of a  $\vmax = 163 \kms$, $\Mvir \simeq 10^{12} \Msun$, 
Milky-Way type halo:
\beq
\T6 \equiv \frac{T}{10^6 \rm{K}}, \quad \t9 \equiv \frac{t_f}{8 \rm{Gyr}}.
\eeq
Here, $8$ Gyr is the time since 
a halo of this temperature
has grown by roughly a factor of 2
according to equation \ref{eq:mass_growth}.

\begin{figure} %fig2
\centering 
\vspace{0pt} 
\epsfig{file=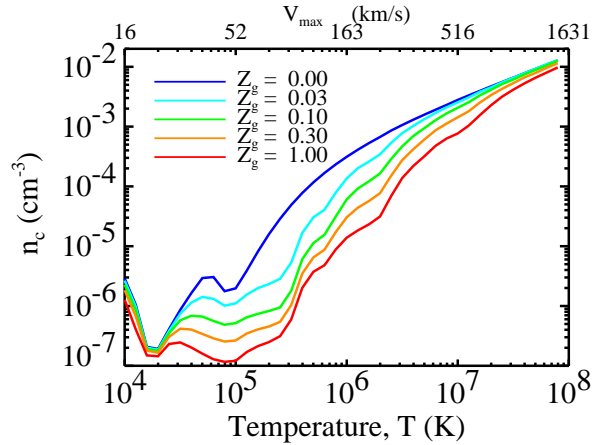,width=\linewidth} 
\vspace{0pt}
\caption{
The cooling density as a function of halo temperature 
for several choices of gas metallicity.  The cooling density
increases dramatically with halo temperature.  Thus massive
halos can retain a larger fraction of their baryons in the
form of hot gas then low mass halos.
}\label{fig:ncool}
\end{figure}

We stress that the temperature scaling  in equation \ref{eq:ncool} and
all of  the analytic expressions  that  follow are only  accurate  for
galaxy  size halos.  A more accurate calculation,
 is shown
in Figure \ref{fig:ncool} for several assumed
gas metallicities.  We have used the mass-doubling
time in equation 
\ref{eq:mass_growth} to set the halo formation time at each
temperature.  This
figure, and all of the  figures to follow, 
rely on the tabulated cooling functions shown in
Figure \ref{fig:coolfunc}.

\subsection{The cooling radius}
  
\citet{wf:91} applied the concept of the cooling density as
a function of radius in a dark matter halo, and introduced
the concept of a cooling radius  as a method to
track the amount of cold gas available to form stars.  This
method has been adopted by most subsequent semi-analytic
models of galaxy formation and seems to do an adequate
job of reproducing the results of hydrodynamic simulations.
The cooling radius, $\rcool$, is defined as the
radius where the cooling density
matches the initial hot gas density: 
\beq
\label{eq:rcooldef}
\rho_{\rm g}^i(\rcool) \equiv \rhocool,
\eeq
unless $\rhocool < \rho_{\rm g}^i(\rvir)$, in which case we will
define $\rcool=\rvir$.
For example, if the 
initial hot gas density follows that of a singular isothermal
sphere then the cooling radius is
\beq
\label{eq:rcoolsis}
\rcool^{^{\rm SIS}} = \sqrt{{\fb \vmax^2}\over{4 \pi G \rhocool}} 
\simeq \, 217 \, {\rm kpc} \quad \T6^{-1/2} \Lt^{1/2},  
\eeq
for  $\rhocool > \rho_{\rm g}^i(\rvir) $.

\begin{figure}
\centering 
\vspace{0pt} 
\epsfig{file=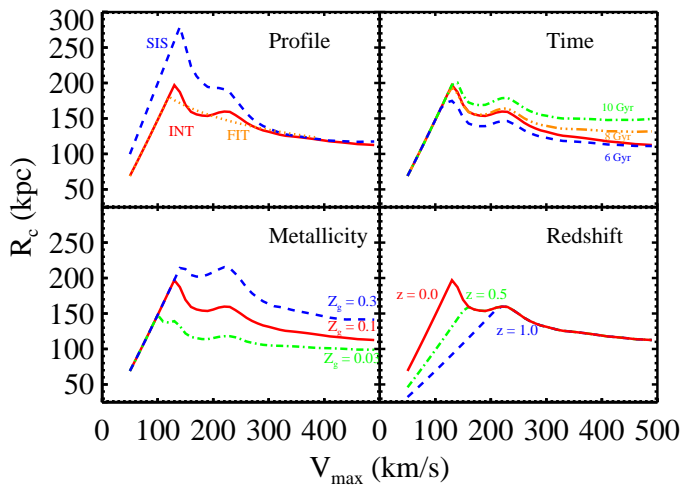,width=\linewidth} 
\vspace{5pt}
\caption{Cooling radius, $\rcool$, as a function of halo $\vmax$.
In all panels, 
the solid line is $\rcool$ calculated with our 
fiducial choices of initial hot profile
(equation~\protect{\ref{eq:init}}), metallicity ($Z_g=0.1$) and
formation time (equation~\protect{\ref{eq:mass_growth}}) at $z=0$
as a function of halo $\vmax$ (assuming an NFW halo). 
Upper Left: Dependence of $\rcool$ on the initial hot gas
profile.  The short-dashed line assumes an SIS gas distribution
and halo profile.  The ``INT'' label refers to our
initial hot gas profile, and the dotted
line is our power-law approximation to this result
(equation~\protect{\ref{eq:rcoolfit}}).  Note that an
SIS halo has a larger virial radius at fixed $\vmax$.
Upper Right: Dependence of $\rcool$ on formation time.
Lower Right: Dependence of $\rcool$ on redshift, for a fixed formation
time (changing only the $\vmax$ to $\rvir$ 
relation).
Lower Left: Dependence of $\rcool$ on metallicity.
}\label{fig:rcool}
\end{figure}

While simple, the isothermal assumption
is not a very good approximation for what we
expect the initial hot 
gas profile to look like (Figure \ref{fig:profile})
and this can lead to large errors in the estimated
 size of $\rcool$ (see upper left panel of Figure \ref{fig:rcool}). 
If instead we use our adopted initial gas profile (equation \ref{eq:init})
then the value of $\rcool$ can be determined by solving a simple
cubic equation.
We find that an approximate fit to the solution for galaxy-size 
halos is
\beq
\label{eq:rcoolfit}
\rcool \simeq 157 \, {\rm kpc} \, \quad \T6^{-1/8} \Lt^{1/3}.
\eeq
This expression is a good fit for the exact cooling
radius solution for halos with $\vmax =\simeq 120 -400 \kms$,
as shown by the dotted line in the upper-left panel of
Figure \ref{fig:rcool} (see below).
For $\vmax \lsim 120 \kms$, the 
virial
radius will be the relevant outer radius (long dash line in Fig. \ref{fig:rcool}):
\beq
\label{eq:rvir}
\rvir \simeq 253 \, {\rm kpc} \, \quad \T6^{0.55}.
\eeq
The above expressions are valid for $z=0$ in our cosmology.

Figure \ref{fig:rcool} shows the exact solution to equations \ref{eq:rhocool} and \ref{eq:rcooldef} for various assumptions for the initial hot gas profile, 
halo formation time, redshift, and  gas metallicity (clockwise from
upper left) as a function of halo $\vmax$.
The solid line in each panel shows the derived cooling radius
for our fiducial set of assumptions, and 
the long-dashed line in each panel shows the halo virial radius.
As seen in the bottom left panel, 
gas metallicity is important in setting the cooling radius
because metal rich gas 
can cool more efficiently (and therefore at lower density and larger
radius) than metal poor gas (see Fig \ref{fig:ncool}).  
Note in the upper left panel that there
is a large difference between assuming an SIS initial 
gas profile (short dash) and our NFW-inspired assumption (solid). 
Not only do NFW halos
fall off more quickly in density at large radius, but they have
smaller virial radii for a fixed $\vmax$.

\section{A two-phase model of cooling}
\label{sec:two-phase}

In the  standard prescription, the evolution of  $\rcool$ with time is
used to evaluate  the amount of  gas available to  form stars.  All of
the gas within the $\rcool$ sphere cools into the central galaxy (over
the halo formation time).  The gas outside of this sphere is assumed to 
stay there, at  the virial temperature of the  halo, tracing the 
background dark halo profile. As time goes on, the cooling radius grows,  
and so does the supply of cold, star-forming gas.    
In  essence,  $\rcool$   is   used as  a
book-keeping tool, since clearly  this shell-like structure of hot gas
represents an unphysical, hydro-dynamically unstable configuration (at
least if $\rcool < \rvir$).  The physical situation this approximation
most closely mirrors is one in which all of the gas within the cooling
radius  cools and contracts monolithically
over the  cooling time,  with hot gas from
the outer regions moving in as a result.  The implicit assumption
is that the thermal instability inherent in the gas is unimportant
in governing the gas infall within the cooling radius.

A different, perhaps more physically-motivated  picture   arises by
considering the two-phase nature of the gas.  As mentioned, gas within
$\rcool$ is subject to the thermal instability, and  will tend to cool
via cloud fragmentation.  That is, not all of  the gas within $\rcool$
will cool, but rather a two-phase (warm/hot) medium will develop.  
Warm ($\sim   10^4$K) clouds will  form and  grow until the background
density of hot  gas is reduced  to roughly $\rhocool$.  Thus there  is
always a  core of hot gas  that extends to the  center of the halo and
that can provide pressure support for the hot gas outside of $\rcool$.

\subsection{Cloud Formation and the Thermal Instability}
\label{ssec:ti}

\citet{field:65} first studied the thermal stability of astrophysical gases,
which was extended to non-equilibrium systems by \citet{balbus:86}.  In the 
equilibrium case with no heating the instability criteria is 
\beq
\left | {{\partial \ln{\Lambda}}\over{\partial \ln{T}}} \right |_P < 1.
\eeq
Using the fitting formula described in Appendix \ref{sec:coolfunc}
(equation~\ref{eq:coolfunc}), we see that hot gas should be unstable 
in all galaxy-size systems.  Specifically, gas will tend
to fragment in halos with temperatures
above the metal-line cooling temperature $T_{m}\simeq 1.5\times 10^5$K.
The range of instability extends below
 $T_{m}$ if the gas metallicity is 
less than solar.  

The thermal instability leads to the rapid growth of perturbations
and to the formation of warm gas fragments within the hot gas background
\citep[see \S\ref{ssec:frag} and also][]{ml:90}.
Perturbations can be either in the gas temperature or density
and may be seeded by the accretion of substructure into the galaxy's halo.
Most of the perturbing halos will have virial temperatures much
below $10^4$K and therefore will not gravitationally bind the
clouds that form.
Of course, some of the most massive subhalos may drive perturbations to  
become gravitationally bound to them; however, as we will be focusing 
on a scenario with an order of magnitude more clouds than massive
dark matter substructures,
we will assume that this is not the case for
most of the clouds. For a treatment of warm clouds
in dark matter substructure see \citet{gs:04} and \citet{smw:02}.

The overdense, low-temperature regions that fragment  will  cool via
atomic line cooling until  they reach a temperature  of $\Tw = 10^4$ K
and   form clouds embedded within    the hot, high-pressure background
medium.  Further   cooling of the warm  gas  into cold  ($\sim 300$K)
material likely will be prevented because of the presence of the 
extragalactic ionizing background, at least for the typical cloud 
densities that we derive below.

\begin{figure}
\centering 
\vspace{0pt} 
\epsfig{file=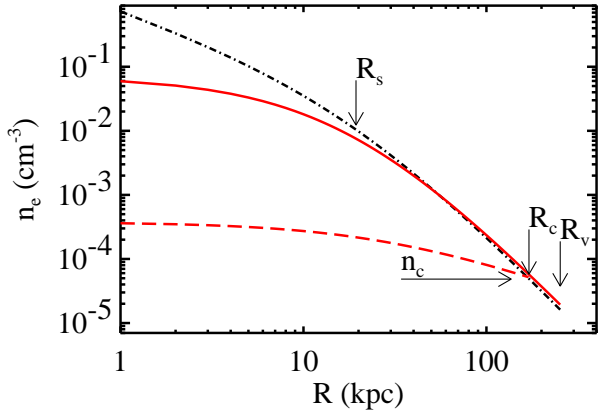,width=\linewidth} 
\vspace{0pt}
\caption{The solid line shows the initial gas profile given by 
equation~\protect{\ref{eq:init}} for a halo with $T=10^6$K 
($\vmax = 163 \kms$).  
For comparison, an NFW profile (dot-dashed line) 
is shown normalized to the same total mass.
The virial radius $\rvir$, the NFW scale radius $\rs$, and the cooling
radius $\rcool$ are also marked.  In the standard cooling argument, all gas 
within the cooling radius contracts to form a central galaxy.  In 
our two-phase 
model, there is a core of hot gas (dashed line) that extends to the center 
of the halo providing pressure support for the gas. In our scenario,
the mass that cools is the integrated mass difference
between the solid line and the dashed line.
}\label{fig:profile}
\end{figure}

\subsection{Residual Hot Profile}
After the clouds form out of the original hot gas halo,
a residual hot gas component will be left.
We can work out a model for this distribution
by assuming that the gas returns to hydrostatic equilibrium within
the gravitational potential of its dark matter halo.
If the residual hot gas does not radiate significantly then it will 
adjust to the pressure change adiabatically.  This is roughly what 
is seen in the cores of the clusters simulated 
\cite[][]{frenk:99} using non-radiative codes 
(note especially the high-resolution result of Bryant's code). Thus
we assume that the gas is adiabatic within $\rcool$, with  $P
\propto \rho_h^{5/3}$  \citep[as adopted by ][]{mm:96}.
Of course, it would be useful to test this assumption
with more detailed multi-phase cooling simulations in the future.

If we normalize by demanding that the hot gas reaches the cooling density 
at the cooling radius, and assume an NFW gravitational potential
(neglecting the contribution of the baryons) then we find 
that the temperature and density profiles of the 
residual hot gas halo follow
(see Appendix  \ref{sec:hydrostatic})
\beqa
\label{eq:ap}
\rho_{\rm h}(x) & = & \rho_c \lb 1 + \frac{3.7}{x}\ln(1+x) - 
\frac{3.7}{\ccool}\ln(1+\ccool) \rb^{3/2}  \\ \nonumber
T_{\rm h}(x) & = & T \lb 1 + \frac{3.7}{x}\ln(1+x) - 
\frac{3.7}{\ccool}\ln(1+\ccool) \rb, 
\eeqa
where the radius $R$ is expressed as $x \equiv R/\rs$ and
$\ccool \equiv \rcool/\rs$.  We have assumed that the hot gas
temperature at $\rcool$ is equal to the halo temperature $T$,
defined with respect  to $\vmax$ in equation \ref{eq:temp}.   
If $\rcool < \rvir$, we assume that the profile outside
of $\rcool$ is isothermal.\footnote{
One may worry that this solution gives values of the hot
gas density that are slightly higher than the cooling density
at small radius.  However, this is only
true if this gas was sitting at this density for a time $t_f$ with the
same temperature.  As discussed in Appendix \ref{sec:hydrostatic},
the gas in the core has likely fallen into the halo center 
more recently than $t_f$, and was heated adiabatically as it fell.
The higher gas temperature and the shorter time available for
cooling will act to increase the cooling density 
of the central gas, and allow it to exist as hot material at a 
higher density than
the global cooling density of the halo.}  
This solution is plotted for a halo with $\vmax=163 \kms$ in Figure 
\ref{fig:profile}. 
The presence of a hot gas core implies that at least some fraction 
of the gas within the cooling radius remains hot.

For simplicity in the calculations that follow, we
work  under   the approximation that   the  temperature,  density, and
pressure of the hot  gas can be  treated as constants as a function of
radius within $\rcool$.  As can be seen in Figure \ref{fig:profile}, 
this is certainly a reasonable approximation for the density.  
In Appendix \ref{sec:hydrostatic} we show that the volume-averaged 
temperature, density, and pressure of the hot gas within the cooling 
radius for expression (\ref{eq:ap}) are given by
\beqa
\label{eq:eta}
\bar{T_h} = \eta_T T, \quad \bar{\rho_h} = \etarho \rho_c,  \quad
\bar{P_h} =  \etap P_c, 
\eeqa
with $\eta_T \simeq 1.0,$ $\eta_{\rm d} \simeq 1.35$, 
and $\eta_P  \simeq   2.7$.
We adopt these values in our treatment below.

\begin{figure} %fig5
\centering 
\vspace{0pt} 
\epsfig{file=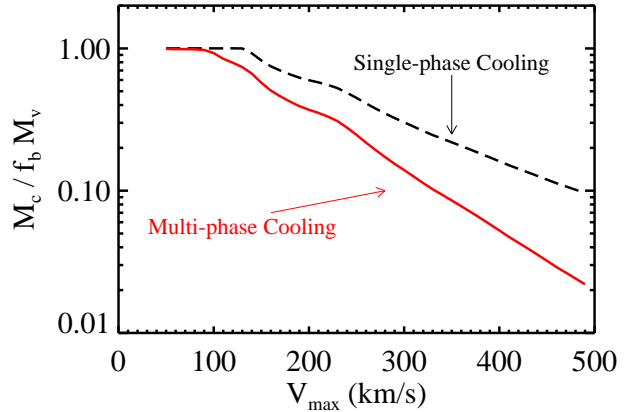,width=\linewidth} 
\vspace{0pt}
\caption{The fraction of baryons in the halo that cool
as a function of
halo maximum circular velocity in our model 
(solid line).  This fraction quickly 
falls from $\sim 90\%$ at $100 \, \kms$ to  $\sim 2\%$ at 
$500 \, \kms$.
Also shown is the same quantity computed using 
the standard, single phase cooling scenario 
(dashed line). The difference arises from the additional hot gas core
that develops in the multi-phase treatment.
}\label{fig:diskfrac}
\end{figure}

\subsection{Cooling Efficiency}

Interestingly, including the simple expectation of a hot
gas core  changes
the cooled-gas fraction in galaxy-size halos appreciably compared
to the standard estimate.
In the standard model, all of the gas inside of the cooling
radius is assumed to cool.  For the initial gas profile given in
equation \ref{eq:init} this mass is
\beq
M_{\rcool} \equiv M_g^i(< \rcool) = M_b \frac{g(\ccool)}{g(\cvir)},
\eeq
where $M_b = f_b \Mvir$.
In our picture, 
$\massrcool$ is divided between cooled gas and the hot gas corona.
The total mass in cooled
material, $\masscool$, is always less than $\massrcool$
because of the presence of a hot core of mass $\masshot$:
\beq
\label{eq:mcool}
\masscool= M_{\rcool} - \masshot, \quad \masshot = \frac{4}{3}\pi \bar{\rho_{\rm h}} \rcool^3.
\eeq
Here $\bar{\rho_h} \propto \rhocool$ is the average density of the residual
hot gas profile within $\rcool$. 

The difference is illustrated explicitly in 
Figure  \ref{fig:diskfrac}.  Shown is
 the  fraction of  baryons  that have
cooled in the halo for both the  single phase (dashed) and multi-phase
cooling (solid) as a function of halo $\vmax$.  For galaxy-size
halos, the   multi-phase treatment reduces   the cold gas  fraction by
$\sim 40\%$ compared to the standard case.  
This  difference will be
amplified if some fraction of  the the gas that cools persists in the
halo as warm clouds (see \S~\ref{sec:infall}). 
The effect of the hot core is
more important for high-mass halos because the cooling density 
increases with temperature (see Fig. \ref{fig:ncool}).
For $\vmax \simeq 500 \kms$ systems, 
the total amount of cooled gas is reduced by
a factor of $\sim 5$ compared to the standard treatment.  As discussed
in \S \ref{sec:lf}, this may have important implications for understanding
the bright cutoff in the galaxy luminosity function.

\subsection{Cloud Size and Density}

For cloud masses of interest, self-gravity
will not be important in setting cloud sizes 
(see \S\ref{ssec:jm}).
Instead, pressure-confinement will set a typical cloud
pressure and density.  The implied density of 
a cloud is
\beqa
\label{eq:rhow}
\rhow = \rhocool \frac{\etap \Th}{\Tw},
\eeqa
where $\Tw$ is the temperature of the warm cloud.
We assume that the clouds are roughly constant density,
so that a cloud of mass 
$\mcloud$ will have a characteristic radius
\beq
\label{eq:rcl}
\rcl =
\lb {{3 \mcloud}\over{4 \pi \rhow}} \rb^{1/3}
\simeq 0.8 {\rm kpc} \quad \m6^{1/3} \T6^{-1} \Lt^{1/3},
\eeq
where $\m6 = \mcloud/10^6 \Msun$, 
and we have used $\etap = 2.7$ and $\Tw = 10^4K$.  

At this stage, the cloud mass is the primary unknown parameter. 
We have normalized our cloud size using a characteristic
mass $\mcloud = 10^6 \Msun$, and we argue below that this may
be an suitable mass for a variety of reasons.  Modeling the
underlying mechanisms that determine
cloud masses  is beyond the scope of the current work.  
We will instead attempt to constrain the allowed parameter space of
clouds using both theoretical limits and, later, observational hints.
Specifically, in the next section we consider various processes in
the halo that will act to destroy clouds, and use these to set limits
on viable cloud masses.  

\begin{figure}
\centering 
\vspace{0pt} 
\epsfig{file=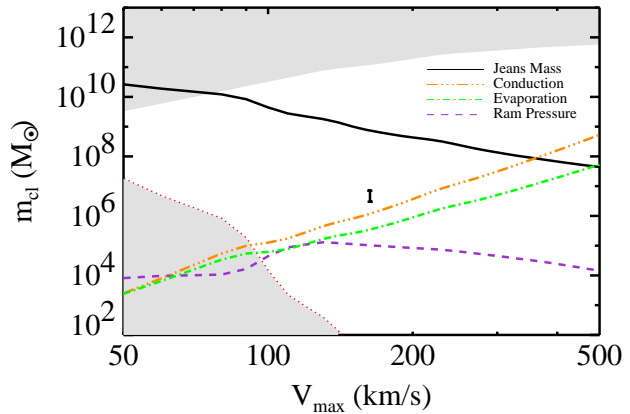,width=\linewidth} 
\vspace{0pt}
\caption{Properties of warm clouds as a function of $\vmax$. 
The upper shaded region shows the total mass within $\rcool$ as a function
of the halo's maximum circular velocity.  Clearly the cloud mass must be 
less than this.  The solid line shows the Jeans mass of the cloud. The
triple-dot-dashed line shows the minimum mass cloud that
can form in the presence of conduction with $f_s=0.2$
(see \S \ref{ssec:conduct}). Masses above the dot-dashed will survive
conductive evaporation over a time $t_f$ (again taking $f_s=0.2$).
For masses below the short-dashed line, ram pressure drag will cause clouds
to move more slowly than the halo velocity, making them
unlikely candidates for High Velocity Clouds.
The shaded region in the lower left
shows where clouds will be destroyed by Kelvin-Helmholtz
instability by having a cooling time longer than the cloud sound
crossing time.  The error bar shows the cloud masses  of interest for 
Milky-Way size  halos ($\vmax = 163 \kms$) 
which help  explain HVCs,  high-ion absorption
systems and the Milky-Way  galaxy mass within  our picture
(see \S \ref{sec:hvc}, \S \ref{sec:abs}, and \S \ref{ssec:total} 
respectively).  
}\label{fig:masses}
\end{figure}

\section{Cloud masses}
\label{sec:cloud-masses}
The physical processes that can act to limit the masses of warm clouds
include  conduction,  evaporation, the Kelvin-Helmholtz  instability, 
and the speed at which clouds can cool.
Except for the last case, these processes impose a
lower-mass    limit on clouds   that  can   survive.   If the  initial
fluctuation distribution  is  power-law, it  is  perhaps reasonable to
assume that clouds will tend to inhabit the lowest-mass regime allowed
\citep{lm:92}.   Another interesting mass is the Jeans mass,
which does not necessarily    affect cloud survival,  but may
determine the mass scale above  which star formation becomes efficient.
Similarly, the relative importance of
pressure drag on a cloud compared to the gravitational force will
vary as a function of mass and this will set a lower limit on
mass scales of interest for HVCs.

The discussion that follows is somewhat lengthy, and we provide
a summary now, in conjunction with Figure \ref{fig:masses}, aimed at
the reader who wishes to move beyond this section to the results.
Figure \ref{fig:masses} shows a space of cloud  masses $\mcloud$
versus halo $\vmax$.  In order  to guide the eye, we  have placed an
error bar on the figure to illustrate the cloud masses  of interest for 
Milky-Way size  halos ($\vmax = 163 \kms$) 
which help  explain HVCs,  high-ion absorption
systems and the Milky-Way  galaxy mass within  our picture
(see \S \ref{sec:hvc}, \S \ref{sec:abs}, and \S \ref{ssec:total} 
respectively).  
The upper shaded  region  is excluded on  physical grounds,  
as  masses above its
lower edge exceed the total baryonic mass within the halo's cooling
radius, $\massrcool$.  The shaded region in the lower left corner,
below the  dotted line, is 
excluded because clouds in this  mass-velocity regime will be 
destroyed by Kelvin-Helmholtz instabilities (\S \ref{ssec:khi}). 
The triple-dot-dashed line 
that runs just below the error bar is the characteristic cloud mass that
arises if conduction sets the cloud fragmentation scale in the initial
hot gas halo (\S \ref{ssec:conduct}).  The dot-dashed line shows
the minimum cloud mass that could have survived evaporation 
within the hot gas halo (\S \ref{ssec:evap}).
The solid line is the dividing line  between Jeans stable (below) and
unstable  clouds (\S \ref{ssec:jm}).  
Finally, the dashed line shows the mass below
which ram pressure drag will cause clouds to move at speeds below
$\sim  \vmax$, and  thus be unlikely candidates for HVCs 
(\S \ref{ssec:ram}).  The main conclusion here is that the cloud 
masses of observational interest are viable based on these considerations.

The analytic expressions that follow were calculated assuming
the cooling curve power-law in equation \ref{eq:Lfit}, 
and therefore are valid only for galaxy-size
halos ($\vmax \simeq 60 - 300 \kms$).  The lines in Figure \ref{fig:masses}
were determined using the $Z_g=0.1$ cooling curve shown in Figure \ref{fig:coolfunc}.

\subsection{Cloud Formation: The Ability to Fragment}
\label{ssec:frag}
As a result of the cooling instability discussed  in \S \ref{ssec:ti}, 
the contrast
between  temperature or density fluctuations in the initial
hot halo will begin to grow as  cooling proceeds.  As slightly cooler
regions begin to cool, 
they get denser and in turn, cool even more quickly, and this can
lead to cloud formation.   Specifically, if the over-cool  
region compresses more quickly than the
background medium can cool,  a separate warm cloud will form 
within the hot gas background.  \citet{bl:00}
studied this process in some detail,  and showed that warm,
dense fragments will emerge in the hot medium as long as
the density growth becomes
nonlinear before the cooling   becomes isochoric.  
The condition   for  cloud  formation  is  that  the
sound-crossing time, $\tau_{\lambda} \simeq \lambda_i/c_{\rm h}$,
along  a  perturbation   of wavelength $\lambda_i$,  should be  
less than the characteristic  cooling time for the halo, which
by our definition of the cooling density equals $t_f$.
Here we have introduced $c_{\rm h}=\vmax/\sqrt{2}$ as the
sound speed of the hot gas.  Let us write the eventual cloud mass 
in terms of the initial fluctuation size as
$\mcloud = 4 \pi (\lambda_i/2)^3 \rho_c/3$.
The condition $\tau_{\lambda} < t_f$
sets an upper limit on the cloud masses that will form
\beq
\mcloud \, \lsim 8.4 \times 10^{11} \Msun \T6^{7/3} \Lz^{-1} \t9^2.
\eeq
We conclude that all mass scales of interest should
be able to form  clouds before isochoric cooling occurs.  Indeed, this upper
limit generally exceeds the total baryonic mass available within halos.

\subsection{Cloud Formation: The Conduction Limit}
\label{ssec:conduct}

A more interesting limit arises from considering conduction.
If conduction is important in the hot gas halo, 
this can dampen temperature fluctuations and inhibit the formation
of clouds.
The length scale below which conduction
will be important compared to cooling (or heating) 
is known as the Field length \citep{mb:90,field:65}
\beq
\label{eq:field_length}
\lambda_F = \lb \frac{T \kappa(T)}{n_e^2 \Lambda(T)} \rb^{1/2},
\eeq
where $\kappa$ is the conductivity of the gas.
One can characterize the conductivity as a fraction $f_s < 1$ of the
classical \citet{spit:62} conductivity:
\beq
\kappa \equiv f_s \kappa_{\rm sp} = f_s {{1.84 \times 10^{10} \T6^{5/2}\over
\ln \Lambda_{\rm C}}} {\rm erg} \, \rm{cm}^{-1} \, \rm{s}^{-1} \, K^{-1},
\eeq
where $\ln \Lambda_{\rm C} $ is the Coulomb logarithm, and
we adopt $\ln \Lambda_{\rm C} = 35$ as an appropriate value
for the temperature and density range of interest \citep{cm:77}.
For an unmagnetized plasma, $f_s$ is unity and conduction is
efficient.  The presence of magnetic fields
can make $f_s$ quite small, with $\sim 0.001$ if the fields are uniform 
or moderately tangled \citep{cc:98}. However \citet{nm:01} have shown that 
$f_s \sim  0.1$ in a medium where magnetic fields are chaotic over a wide 
range of length scales.  The results of \citet{zn:03}
imply that $f_s \simeq 0.2$ can help solve the cooling flow
problem in  clusters \citep[see also][]{kn:03}.  
We will adopt $f_s = 0.2$ as our
fiducial value here.  With this choice
we find that the Field length of a hot gas at the cooling density is
\beq
\lambda_F \simeq 11 {\rm kpc} \quad \T6^{1/4} \Lz^{1/2} \t9 f_{0.2}^{1/2}. 
\eeq
Scales smaller than $\lambda_F$ will tend to have a
uniform temperature, and this implies
a characteristic lower-limit on the mass:
$m_{cl}^{\rm F} \equiv 4 \pi (\lambda_F/2)^3 \rho_c/3$.  Using 
typical numbers we find
\beq
\label{eq:fieldmass}
m_{cl}^{\rm F} \simeq 1.2 \times 10^6 \Msun 
\quad  \T6^{11/4} \Lz^{1/2} \t9^2 f_{0.2}^{3/2}.
\eeq
We plot the Field mass  as a function of $\vmax$ for 
$f_s=0.2$ as the triple-dot-dashed line in Figure \ref{fig:masses}.
Note that $m_{cl}^{\rm F}$ as defined above scales as $f_s^{3/2}$.
If conduction operates at $\sim 20\%$ the Spitzer value,
the characteristic mass scale is quite similar to our mass scale
of interest.  

We point out that if $\lambda_F \gsim \rcool$, we 
expect cloud fragmentation to be stabilized completely.
When this occurs, conductive heating from
outside of $\rcool$ can play an important role
in setting the temperature structure of halos.
This occurs when $T \gsim 3.2 \times 10^{7} K$ ($\vmax \gsim 920 \kms$), 
or in massive cluster-size systems.  (Note that the scaling 
in equation \ref{eq:fieldmass} is only valid for $60 \kms \lsim \vmax \lsim 300 \kms$ because
we have assumed a power-law form for the cooling function in 
its derivation.)

\subsection{Cloud Survival: The Kelvin-Helmholtz Instability}
\label{ssec:khi}

Once clouds form they are subject to shearing stresses across their
boundary as they travel through the hot medium.  The flow can
be subject to perturbations, and this is characterized
as the Kelvin-Helmholtz instability (KHI).

The dominate destructive  process is  the
``champagne effect'', which results from the development of a 
low-pressure, fast flow
around the head of the moving cloud \citep{dz:81,murr:93,vfm:97}.  
By Bernoulli's theorem, the pressure exerted by the fast wind 
at the head of the cloud is low, so the
cloud's inner pressure can cause its material to
be pushed out from the top.
\citet{vfm:97} showed that the champagne effect is
stabilized if the cooling  time of a cloud is  shorter than the sound 
crossing time of the cloud.  That is,  if the pressure  waves
 inside the cloud are damped
by radiative cooling before they can cross the cloud, then the inner
part of the cloud  cannot respond to  produce the over-spilling. 
Note that this result holds even for clouds that are
in thermal equilibrium 
with a  background field, as assumed here.

Based on the work of \citet{vfm:97}, we would
like to compare the cooling time of our clouds to the sound crossing time.
This will determine if they are stable against the KHI.
The sound crossing time of the cloud is $\tau_{\lambda}^{\rm cl} =
\rcl/c_{\rm w}$, where $c_{\rm w} \simeq 11.5 \kms$  
is the speed of sound in the warm medium. 
Using equation \ref{eq:rcl} for the cloud size we obtain
\beq 
\tau_{\lambda}^{\rm cl} \simeq 6.7 \times 10^{7} {\rm yrs} 
\quad m_6^{1/3} \T6^{-1} \Lt^{1/3}. 
\eeq
We compare this to the
the cooling time of a cloud of temperature $T_{\rm w}$ and density
$\rhow$:
\beq
\tau_{\rm c}^{\rm cl} = \frac{3 \mu_e^2 m_p k_b \Tw}{2 \mup \rhow \Lambda(\Tw)}
\simeq 1.6\times 10^{6} {\rm yrs} \quad \T6^{-3} \Lz\t9,
\eeq
where we have used 
$\Lambda(T_w=10^4K) = 4.9 \times 10^{-24} {\rm cm}^{3} {\rm erg} \,{\rm s}^{-1}$ (see Figure \ref{fig:coolfunc}).
This allows us to define a characteristic KHI mass, above which clouds 
will be stable.  By setting $\tau_{\rcl} (\mcloud) = \tau_{\rm c}$ we
obtain
\beq
\mcloud^{\rm KHI} \simeq 10.5 \Msun 
\quad \T6^{-6} \Lt^2,
\eeq
and note that the mass above which clouds are stable
is a very strong function of the hot gas temperature.  This
is seen clearly by the shaded region in the lower left of 
Figure \ref{fig:masses}.  
As the host halo's temperature goes down, clouds become
less dense (see equation \ref{eq:rhow}), their cooling
times increase, and they are more susceptible to the KHI.
We see that cooling alone stabilizes most cloud masses of interest
except  in low-temperature halos.\footnote{Note that even in 
low-temperature halos, 
magnetic effects may stabilize clouds against the KHI
\citep{chan:61,miura:84,mbr:96}.}

\subsection{Cloud Survival: Conductive Evaporation}
\label{ssec:evap}

Clouds may also be evaporated by conduction from the 
surrounding hot gas.  The characteristic evaporation time scale is 
given by 
\beqa
\label{eq:evap}
\tau_{\rm evap} & =& {{25 k_b \mcloud}\over{16 \pi \mup m_p \kappa(T) \rcl}} \\ \nonumber
&\simeq& 16 {\rm Gyr} \quad \m6^{2/3} \T6^{-3/2} (\Lz\t9)^{-1/3},
\eeqa
\citep{cm:77} where we have taken $f_s = 0.2$.  If we set this equal to the 
halo formation time (i.e. if require that clouds forming at $t_f$ have not 
evaporated by today) this gives us a lower bound on the cloud mass of 
\beqa
\mcloud^{\rm Evap} &=& \sqrt{{3 \pi^2}\over{{4\rhow}}}  
\lb{{16 \mup m_p \kappa(T) t_f}\over{25 k_b}}\rb^{3/2} \\  \nonumber
&\simeq&  3.5 \times 10^5 \Msun \quad \T6^{9/4} \Lz^{1/2} \t9^{2}.
\eeqa
This is shown as the dot-dashed line in Fig. \ref{fig:masses}.  Cloud masses
above this line will not evaporate over a time $t_f$. 

\subsection{Cloud Motion: Ambient Drag}
\label{ssec:ram}

As the cloud moves through the hot gas halo at speed $\vcl$, it will
experience a ``ram pressure'' drag force that opposes its motion
\citep[e.g.][]{ll}
\beq
\label{eq:ram}
F_{\rm ram} =  \frac{1}{2} C_{\rm d} \bar{\rho_h} \vcl^2 \pi \rcl^2,
\eeq
where $C_{\rm d} $ is the drag coefficient.
Clouds reach terminal velocity, $v_{\rm t}$, when the
gravitation force on the cloud is balanced by
ram pressure force:
\beq
v_{\rm t}^2 = 2 {{G M(D) \mcloud}\over{D^2\pi \rcl^2 \bar{\rho_h} C_{\rm d}}} 
=  \vmax^2 \lb \frac{8 \rcl \rho_w}{3 D C_{\rm d} \bar{\rho_h}} \rb.
\eeq
For simplicity, we have assumed that the host halo
is an isothermal sphere, with $D$ the 
distance from the cloud to the halo center.  If we assume that a 
typical distance is $D \simeq \rcool$ then we obtain:
\beq
\label{eq:ram_vel}
{{v_{\rm t}}\over{\vmax}} = \sqrt{{8 \rcl \etap \Th}\over
{ 3 C_{\rm d} \rcool \etarho \Tw}}
\simeq 1.6 \quad m_6^{1/6} \T6^{1/16} \Lt^{-1/12},
\eeq
where we have used $\etarho = 1.35$ and $C_{\rm d} = 1.0$.
Thus if clouds are sufficiently massive they can travel at a typical speed 
$\vcl \simeq \vmax$ and not experience significant deceleration over 
a dynamical time.  The limiting mass, below which $\vcl < \vmax$ is
\beq
\label{eq:ram_mass}
\mcloud^{\rm ram} \simeq 5.1 \times 10^4 \Msun 
\quad \T6^{-3/8} \Lt^{1/2}.
\eeq
This mass scale is plotted as the dashed line in Figure \ref{fig:masses}.
Clouds very much smaller than this are physically viable, 
but their slow speeds would make them unlikely candidates
for ``high-velocity'' clouds.   We return to the effect that ram
pressure drag will have on cloud motion in \S \ref{ssec:coll}. 

\subsection{Cloud Self-Gravity: Jeans Mass}
\label{ssec:jm}

The Jeans mass estimates when a fluid is unstable to self gravity. 
For a cloud confined by a pressure $P_h$ it is given by \citep{spit:78}
\beq
M_J = {{9 c_{\rm w}^4}\over{5 G^{3/2}P_h^{1/2}}} .
\eeq
Warm clouds with temperature  
$\Tw = 10^4$ K have a sound speed $c_{\rm w} = 11.5 \kms$,
and the minimum
cloud mass that is unstable to self gravity is 
\beq
\mcloud^J \simeq 7.2 \times 10^8 \Msun  
\quad \T6^{11/4} \Lt^{1/2}.
\eeq
The Jeans mass for a warm
cloud as a function of halo maximum circular velocity is plotted 
as the solid line in 
Fig. \ref{fig:masses}.

Our expectation is that clouds will be less massive than the 
Jeans mass in galaxy-size halos and 
therefore their self gravity can be ignored.
However, it is possible that in high-mass halos that clouds will
tend to be more massive.  This might happen, e.g.,
if conduction sets the cloud mass (equation \ref{eq:fieldmass}
and the triple-dot-dashed line in Fig. \ref{fig:masses}).
In this case, the clouds may be above the Jeans mass, self-gravitating,
and perhaps form stars in cluster-size halos.  
These objects would likely be spheroidal
systems, with rather low mass-to-light ratios.  Cluster ``galaxies'' 
of this type would not contain dark matter.

\section{Galaxy Formation via cloud infall}
\label{sec:infall}
The warm clouds are accelerated towards the center of the halo,
but we do not expect  that they will settle there immediately.  
Rather, clouds should have some distribution of energy and angular momentum
that will have to be lost before the clouds merge with the central
galaxy.  For example,  clouds  may take on an
angular momentum distribution similar to that of bulk-averaged regions
seen in dark matter halo N-body simulations  \citep{bull:01b} or
of streaming  motions in hydrodynamic  simulations  \citep{bosch:03}.  
Clouds stripped from merging halos are expected to have
a similar distribution  \citep{md:02}.

For concreteness, we
will  assume  that the  clouds take on  an  energy distribution with a
characteristic velocity equal to the halo velocity, $\vcl =  \vmax$.
Cloud-cloud collisions or ram pressure drag eventually lead to cloud orbital
decay.  In our picture, it is  this  cloud infall that sets  the gas
supply that governs the formation of the central galaxy.  In the following
subsections, we discuss expected infall times,  explore
the implied central galaxy mass and residual cloud population,
and discuss the disruption of clouds as they 
approach the galaxy.  

\subsection{Infall Times}
\label{ssec:coll}

As clouds orbit within the hot gas halo they will 
experience ram pressure drag (equation \ref{eq:ram}).
The continuous drag
will
sap energy from the clouds, and this can eventually lead to
orbital decay.
The timescale for this to occur is given by
\beq
\label{eq:tauram}
\tau_{\rm ram} = {{2 \mcloud}\over{\pi C_{\rm d} \rcl^2 \bar{\rho_h} \vcl}}
\simeq 2.6 {\rm Gyr} \quad  m_6^{1/3} \T6^{-1/2} \Lt^{1/3},
\eeq
if $\vcl = \vmax$  and $C_{\rm d}=1$. 
Thus for $10^6 \Msun$ clouds, only those that 
cooled out of the hot gas more
than $\sim 3$Gyr ago would have begun to sink  to the center of
the halo via ram pressure effects.  If the drag coefficient, $C_{\rm d}$
is less than $1$,
then this is a lower limit on the drag decay time.\footnote{We note that
the Reynolds number for clouds in such a halo, Re$\simeq \rcl \vcl/\nu$
(where $\nu$ is the viscosity), is expected to be quite high,
Re$\sim 10^9$, if conduction sets the viscosity.  In this
case,  the drag coefficient, $C_{\rm d}$, is likely to be less than
unity, even for supersonic flow.}
 We show below
that the mass in clouds that we expect to have fallen in  or
formed in the halo since that time
may be a rather large fraction of the
baryonic content of the Galaxy.

Cloud collisions will also be important in triggering cloud infall.
We will assume, for simplicity, that after a collision, most of
the cloud energy goes into heating the cloud material, and that
this is quickly radiated away.
The remaining, likely merged, system
will have low kinetic energy and will quickly fall in to
contribute to the central galaxy.  Thus the cloud infall time
will scale like the cloud-cloud collision time.

The mean free time between cloud collisions can be written as
$\tau_{cc} \simeq (\ncloud \vcl \sigma_{\rm cl})^{-1}$, where the cloud
cross section is $\sigma_{\rm cl} \simeq \pi r_{\rm cl}^2$
and $\ncloud$ is the number density of clouds.
Note that $\ncloud$
depends on the total mass of warm clouds $\masscl$, 
and that this will change as clouds collide and merge or if
new clouds form.  
If we assign
$\ncloud = 3 \masscl /(\mcloud 4 \pi \rcool^3)$ we can write the 
cloud-cloud collision time as
\beq
\label{eq:taucc}
\tau_{cc}= {{4 \mcloud \rcool^3}\over{3 \masscl \vcl \rcl^2}}
\simeq 2.4 {\rm Gyr} \quad
\quad m_6^{1/3} \T6^{9/8} \Lt^{5/6} M_{2.10}^{-1}.
\eeq
Here we have used $M_{2.10} \equiv \masscl/(2 \times 10^{10} \Msun)$  
as the characteristic mass in clouds that we expect to exist 
in our fiducial halo.  We explain this expectation in more detail in 
\S\ref{ssec:total}.  We stress, however, that $\masscl$ should
vary as a function of halo mass and cloud mass,
so their are additional dependencies in equation \ref{eq:taucc} that
are hidden in this variable.

When the density of clouds is high,
$\tau_{cc}$ is small, and clouds will quickly collide and sink to the 
center.  As the total number of clouds drops,
the cloud infall rate will begin to drop as well.
It is useful to consider the simple scenario where
we start with a number density of clouds $\phi_0$ in a fixed 
volume.  In this case the number density of clouds as a function of time 
obeys $d\phi/dt = \phi(t) \tau^{-1} = \phi^2 (\tau_0 \phi_0)^{-1}$,
where $\tau_0$ is the mean free time initially.
The solution is  $\phi(t) = \phi_0/(1+t/\tau_0)$,
so there will  always be a residual cloud population in any halo,
if cloud collisions set the infall rate.

We mention that 
the shortest timescale over which clouds can fall in to 
the central galaxy is the free-fall time, $\tau_{\rm ff}$.  If we
estimate the cloud free-fall time from the cooling radius
as $\tau_{\rm ff} = \rcool/\vmax$ then we obtain
\beq
\tau_{\rm ff} \simeq 0.94 {\rm Gyr} \quad \T6^{-5/8} \Lt^{1/3}.
\eeq
This expression is accurate for halos with maximum circular velocities 
of $120 \kms \lsim \vmax \lsim 400 \kms$ 
(where $\rcool < \rvir$ sets the lower limit 
and the breakdown in the scaling $\Lambda(T) \propto T^{-1}$ sets the
upper limit).
For all cases that we consider, the free-fall timescale is
shorter than both $\tau_{cc}$ and $\tau_{\rm ram}$.

\subsection{Central Galaxy Mass}
\label{ssec:total}

The total mass within the halo cooling radius, $\massrcool$, is divided between
gas in the hot halo core, $\masshot$, and
gas that has cooled since the halo formation time $M_{\rm c}$
(equation \ref{eq:mcool}).
The cooled gas mass is itself shared between
warm clouds, $\masscl$, and the central galaxy $\massgal$.
The mass budget is then described by
\beqa
\label{eq:massbudget}
M_{\rcool} & = & \masshot + M_{\rm c} \\ \nonumber
M_{\rm c} & = &  \masscl + \massgal.
\eeqa

\begin{figure} %fig6
\centering 
\vspace{0pt} 
\epsfig{file=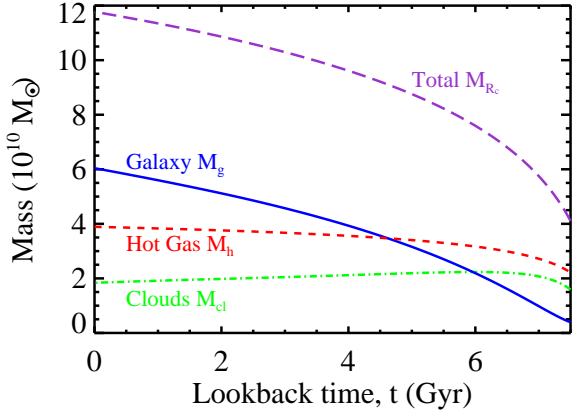,width=\linewidth} 
\vspace{0pt}
\caption{Evolution of each baryonic component as a function of lookback time.
The long dashed line shows the the total baryonic mass within the cooling 
radius, which grows steadily as a function of time.
The central galaxy mass
(solid line) also grows with time, as clouds continue to fall in. 
The total mass in the hot gas core
(short dashed line) remains roughly constant as the cooling radius
grows and cooling density drops.  The total mass
in clouds (dot-dashed line) remains nearly constant, as cloud formation
and infall are balanced.
}
\label{fig:lookback}
\end{figure}

We assume that cooling proceeds by the formation of
clouds and that the infall of clouds leads to galaxy growth.  
The evolution of the total mass in clouds as a function of time
can be modeled as a competition between cold mass accumulation
(as the cooling radius grows) and the rate of cloud ``destruction'' via 
infall onto the galaxy:
\beqa
\label{eq:dmdt}
\frac{d \masscl}{dt} & = &\frac{dM_{\rm c}}{dt} - \frac{d \massgal}{dt}  \\ \nonumber
\frac{d \massgal}{dt} & = & \frac{\masscl}{\tau_{\rm in}}.
\eeqa
We have associated 
the cloud infall rate with the rate of galaxy
growth, and set this equal to $\masscl/\tau_{in}$.
Here $\tau_{in}$ is a characteristic cloud infall time,
chosen to be the minimum of $\tau_{cc}$ and $\tau_{\rm ram}$.
For the fiducial halo and cloud mass discussed in this section,
$\tau_{cc} < \tau_{\rm ram}$, and cloud-cloud collisions dominate 
the infall.

It is   straightforward to solve  this  simple set of equations
(\ref{eq:massbudget}, \ref{eq:dmdt}) in order to evaluate the mass in
each  component. Once we choose a halo $\vmax$, 
the evolution of the hot gas core mass, $\masshot(t)$,
is  governed entirely by  the
evolution of   the cooling radius  and corresponding  evolution in the
cooling density  (equations \ref{eq:ncool}, \ref{eq:rcoolfit},
\ref{eq:mcool}).  The evolution of $M_{\rm c}$ with time can similarly
be determined by  the evolution of $\rcool(t)$ (equation \ref{eq:rcoolfit})  
and $\masshot(t)$ (as just described).  
The other components may be tracked via equation \ref{eq:dmdt}
once one adopts a cloud mass, $\mcloud$, 
and evaluates $\tau_{\rm in}(t)$ using
equation \ref{eq:taucc} or \ref{eq:tauram}.

\begin{figure}
\centering 
\vspace{0pt} 
\epsfig{file=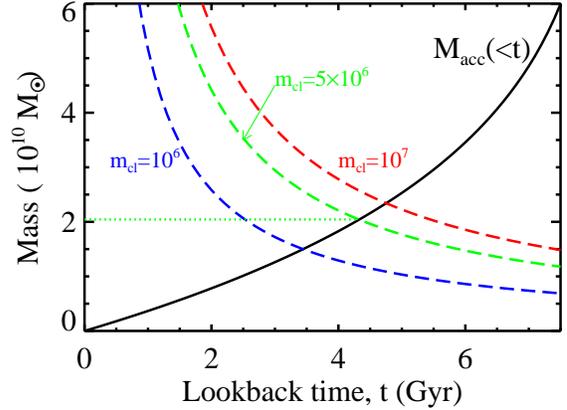,width=\linewidth} 
\vspace{0pt}
\caption{Shown as the solid line is the amount of mass that has cooled 
as a function of lookback time: $M_{\rm acc}(t)=\masscool(0)-\masscool(t)$.
The dashed lines show the total mass in clouds that would
have a collision time, $\tau_{cc}$, 
that is equal to the lookback time plotted on the x-axis.
We show the result for three cloud masses, 
$(1, 5, 10)\times 10^6 \Msun$.  The point where 
a dashed line
intersects the solid line gives the approximate mass in clouds
that can survive until $t=0$ (today).
For a typical cloud mass
$\mcloud=5 \times 10^6 \Msun$, the expected residual mass
in clouds is $M_{\rm cl} \simeq 2 \times 10^{10} \Msun$ (as indicated
by the dotted line), which is 
in good agreement
with what is found by integrating the set of equations 
discussed in the text and shown in Fig. \ref{fig:lookback}. 
}\label{fig:acc}
\end{figure}

Figure \ref{fig:lookback} shows the 
resulting buildup in each mass component as a function of lookback
time for our fiducial ``Milky Way'' halo of $\vmax = 163 \kms$,
and a cloud mass of $\mcloud = 5 \times 10^6 \Msun$.  
The top long-dash line  shows the total baryonic mass within the
cooling radius, $\massrcool$, and the lower
set   of dot-dashed, short-dashed, and solid lines show 
the galactic mass, hot core mass, and total cloud mass ($\massgal$,
$\masshot$, and $\masscl$) respectively.  The most striking
result is that the final galaxy mass, 
$\massgal \simeq 6 \times 10^{10} \Msun$,
is roughly half of the mass it would have been had we adopted 
the standard treatment, and allowed all of the mass within
$\rcool$ to contribute, $\massrcool \simeq 12 \times 10^{10} \Msun$. 
The mass in
clouds peaks rather early at $\masscl \simeq 2 \times 10^{10} \Msun$,
 and then remains relatively constant as
the cloud infall rate is matched by the rate of accumulation of
cooled gas.   The mass within the hot core also remains
relatively constant as a function of lookback time $\masshot \simeq 4 \times 10^{10} \Msun$.
This is because $\masshot \propto \ncool \rcool^3$, and
growth in $\rcool$ at late times is canceled out
by the decrease in $\ncool$ 
(this can be seen via inspection
of the time scalings in equations \ref{eq:ncool}  and \ref{eq:rcoolfit}).

The  final residual cloud mass expected in this scenario is
straightforward to understand without 
having to  solve the differential equations.
Given a  characteristic cloud infall  time
$\tau_{\rm in}$, we expect clouds to remain in the halo as long as
$\tau_{\rm in}$ is longer than the time since the clouds
were formed (or accreted).

Consider a case where $\tau_{\rm in} = \tau_{cc}$ and
cloud-cloud  collisions  dominate infall.  The solid line in Figure
\ref{fig:acc}  shows the amount of mass that has cooled since 
a lookback time $t$.  Specifically,
 $M_{\rm acc}(t) = M_{\rm c}(0) - M_{\rm c}(t)$.  We have
used the same $\vmax = 163 \kms$ halo discussed in conjunction with 
Figure \ref{fig:lookback}.  
The three dashed lines in Figure \ref{fig:acc} correspond to
cloud-cloud collision  times computed  using three   different cloud
masses $\mcloud = (1$, $5$, and $10) \times 10^6 \Msun$.  In each case
we show the amount of mass needed in clouds to get a $\tau_{cc}$ equal
to the time on the x-axis.
Note that  cloud-cloud collision times are short if the total mass in
clouds is large.
The point where the dashed lines 
cross the solid line corresponds to the mass in clouds that could
have survived until the present day without merging into the  galaxy.
Thus, for the central  line  ($\mcloud = 5   \times 10^{6} \Msun$)  we
expect a  final mass in clouds  of $\masscl  \simeq 2 \times 10^{10}
\Msun$ to have survived until the present day.  
>From Figure \ref{fig:lookback} we see that this is very close to the
mass in clouds derived from integrating equations (\ref{eq:massbudget}
and \ref{eq:dmdt}).  This simple method allows a quick way to estimate the
final central galaxy mass: $M_{\rm g}(0) \simeq  M_{\rm c}(0) - M_{\rm
acc}(\tau)$.

\begin{figure}
\centering 
\vspace{0pt} 
\epsfig{file=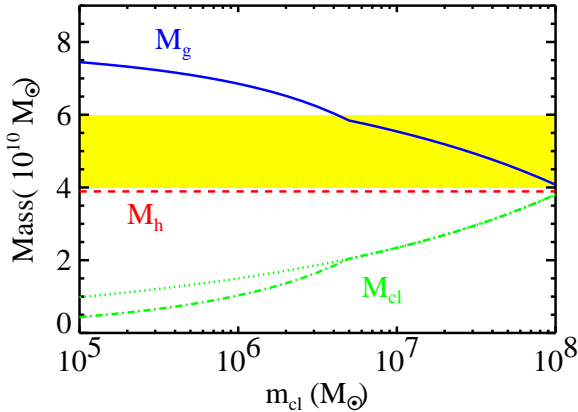,width=\linewidth} 
\vspace{0pt}
\caption{Shown is the mass in the hot gas core (dashed), 
mass in warm clouds, with and without ram pressure (dot-dashed and dotted)
and mass in the galaxy (solid) as a function of the warm cloud mass.
The shaded region shows the range of masses for the Milky
Way \protect{\citep{db:98}}. Cloud masses in the range 
$(4-100) \times 10^{6} \Msun$ allow agreement with estimated
Milky Way mass without including any blow-out feedback.
}\label{fig:clmass}
\end{figure}

We have applied this simple treatment in   Figure \ref{fig:clmass}  in  
order to estimate the range of cloud masses that may help explain the 
baryonic mass of the Milky Way.  The shaded band shows the estimated range of
Milky Way galaxy masses derived by \citet{db:98}.  The solid line shows 
the final galaxy mass, $\massgal$, and  the dashed line shows the residual
total mass in clouds, $\masscl$, as a function of the individual cloud
mass,  $\mcloud$.  Note that  the Milky  Way mass is  matched well for
$\mcloud \simeq 4\times 10^{6} - 10^{8} \Msun$ without including
any blow-out feedback. Note however, that even if cloud masses
are small ($\mcloud  \lsim 10^6 \Msun$) and fall in quickly to
assemble the Galaxy, explaining the mass
of the Milky Way is much easier in this picture because of the
substantial hot gas core.
The kinks in the lines occur at the cloud mass
below which ram pressure dominates
cloud infall.  Cloud-cloud collisions are more important for massive
clouds.  
  The dotted line shows how the mass in clouds would
change had we set the cloud drag constant to zero, so that ram
pressure was unimportant in cloud evolution.

\subsection{Tidal Disruption}
\label{ssec:tidal}
As clouds approach the central galaxy they will experience strong
tidal forces and may be destroyed.   If clouds are broken up
before impacting the galaxy, this will prevent them from
causing too much heating or damaging the disk.
Tidally deformed clouds will tend to be quite large, and perhaps can be
identified with the large HI complexes that are well known
to exist in proximity to the Milky Way \citep{tripp:03,ww:91,blitz:99}.

Clouds will be disrupted when the tidal force from the host
potential overcomes the pressure confinement of the clouds.
At a distance $D$ from the host center, assuming that
the host potential is roughly isothermal, the
tidal force felt across a cloud of radius $r_{\rm cl} \ll D$ 
is approximated as
\beq
F_{T} \simeq {{G M_{\rm dm}(D) m_{\rm cl} r_{\rm cl}}\over{D^3}}.
\eeq
This can be compared to the typical pressure force 
\beq
F_{P} \simeq \pi \rcl^2 \eta_P \rhocool \Th.
\eeq
The radius $R_{\rm d}$ where the two forces are equal defines the surface of a 
sphere of disruption for the clouds.
For an SIS density distribution $M(R)=G^{-1} \vmax^2 R$, the tidal 
sphere for destroying clouds can be derived analytically:
\beq
R_{\rm d} \simeq  13 \rm{kpc} 
\quad m_6^{1/3} \T6^{-1/2} \Lt^{1/3}.
\eeq
In the galaxy mass regime ($100 \kms < \vmax < 350 \kms$) the value of $R_d$
for a SIS and a NFW halo are very similar.
We see that the sphere of disruption is larger than 
the size of the Milky Way's disk.  It is therefore unlikely that
the disk will be heated significantly from impacting clouds.
Interestingly, this distance would
be consistent with distance limits for many of the large HVC
complexes \citep[e.g.][]{woer:99a,woer:99b,wakker:01}.

\section{Residual clouds as HVCs}
\label{sec:hvc}

 Seen in 21 cm HI emission, High Velocity Clouds (HVCs) 
have been studied for more than four decades \citep{mor:63}.
Interpreting the observed properties of HVCs
in terms of physical
parameters requires knowing the radial distance, $D$, from the
Sun.  This is the major 
observational unknown that has fueled the debate over their origin
since their discovery. 

Models for HVCs range from condensed ``Galactic Fountain'' gas at 
$D\simeq 5$kpc \citep{sf:76,breg:80}, to large, 
extra-galactic objects associated with the Local Group $D\simeq 1$Mpc 
\citep{vers:69,arp:85,blitz:99,blitz:02,smw:02,malo:03}.  
In our picture, the HVCs are ``circumgalactic'', within the cooling 
radius of the Galaxy, and bear resemblance to the $D \simeq 100$kpc 
population suggested by \citet{oort:66}.
~\footnote{We mention that a fragmentary, pressure-supported HVC population 
similar to the one we suggest might arise with a Local Group barycenter, 
as long as Andromeda and the Galaxy share a common hot gas halo 
(L. Blitz, private communication).  Here we will focus on the Galaxy 
as an isolated halo as we have throughout this work.}

We  expect that most of the mass in each cloud
is in the form of ionized hydrogen at a temperature
$T \simeq 10^4 K$.   Clouds of this  kind would have a
velocity distribution 
full-width-half-max  (FWHM) of $\Delta v \simeq   27 \kms$.  The line
width distribution of  Compact  HVCs studied by \citet{dbb:02}  has a
median FWHM of $\Delta v = 25 \kms$.  The agreement with predicted and
observed line widths is encouraging.

The recent HIPASS  survey cataloged  HVCs
over the  entire  southern sky \citep{putm:02}.  They find
that the radial velocity distribution of their
clouds is narrow when 
plotted with respect to the Galactic Standard of Rest, with
$\sigma_r = 115
\kms$ (it is $\sigma_r=185 \kms$ for the Local  Standard of Rest).
It peaks\footnote{The HVC velocity distribution
set in the Local Group Standard of
Rest peaks  near $\sim -75 \kms$,
and has about the same  dispersion  as the  Galactic Standard of Rest
distribution.} near $\sim 0 \kms$.  As  discussed  in the introduction,  
completely  disjoint dynamical
models of the  Milky Way lead us to  choose  a fiducial ``Milky  Way''
dark halo with $\vmax = 163 \kms$, and therefore a velocity dispersion
very  close  to   the   GSR distribution of   HVCs:  $\sigma_r  \simeq
\vmax/\sqrt{2} = 115 \kms$.  Since the residual clouds in our scenario
should  roughly  take on  the  dark halo's  velocity distribution, we
expect them to match the observed HVC distribution quite well.

The HIPASS HVC population has a characteristic peak $HI$ column density
of $N_{HI} \simeq 10^{19} \cm2$ and a characteristic 
angular size of $\theta \simeq 0.5$deg$^2$ \citep{putm:02}.
The expected hydrogen space density for an individual cloud 
in our model is
\beq
\label{eq:cldense}
n_{H} = \frac{\rhowarm f_H} {m_p} \simeq
1.4 \times 10^{-2} \rm{cm}^{-2} \quad \T6^3 \Lt^{-1},
\eeq
where we assume that the mass fraction in Hydrogen is $f_H = 0.7$.
The  column density in $HI$ through a cloud can be estimated 
via $N_{HI} = 2 \rcl n_{H} \epsilon_{HI}$, with
$\epsilon_{HI}$ the fraction of neutral hydrogen.
With $\epsilon_{HI}=0.1$ \citep{malo:03} we obtain
\beq
\label{eq:column}
N_{HI} \simeq
6.8 \times 10^{18} \rm{cm}^{-2} \quad m_6^{1/3} \T6^2 \Lt^{-2/3}. 
\eeq
If we demand
$N_{HI} = 10^{19} \rm{cm}^{-2}  $ for our typical cloud in order
to match observations, this would imply 
$\mcloud \simeq 3 \times 10^6 \Msun$.

The angular size of the cloud is related to the cloud radius by
\beq
\tcl \simeq 10^4 \lb{{\rcl}\over{D}}\rb^2 {\rm deg}^2
\eeq
where $D$ is the distance to the cloud. 
If we set $D=\rcool$ (equation \ref{eq:rcoolfit}), 
then the typical cloud area on the sky
will be
\beq
\label{eq:tcl}
\tcl \simeq 0.25 \, {\rm deg}^2 \quad m_6^{2/3} \T6^{-7/4}.
\eeq 
In order to match a  typical size of  $\tcl = 0.5$deg$^2$ we will need
$\mcloud  \simeq 3 \times  10^6 \Msun$,  which  is nicely in line with
what   we needed  to match  the  column   density above.   
Of course, the angular sizes observed correspond to 
$HI$ sizes, and one might expect the outer radius
in neutral material to be somewhat smaller that the
full cloud radius.
In the  models  of \citet{malo:03},
$r_{HI}  \simeq 0.7 \rcl$ for constant-density  clouds similar to the
type we consider here.  If we adopt this assumption, then the coefficient
in equation \ref{eq:tcl}   would scale  to   $0.18$ deg$^2$, and push 
our preferred mass to $\mcloud \simeq 5 \times 10^6 \Msun$.

Finally, there are roughly 2000 HVCs in the HIPASS sample covering the
southern sky.  If we double  this, we can  estimate that the full halo
should  contain  $\sim 4000$  such clouds.   The  number of  clouds we
expect in the halo is simply $N_{\rm cl} \simeq
\masscl/\mcloud$.  In the previous section, we assumed cloud masses
of $\mcloud = 5 \times 10^6 \Msun$ and
computed that the total residual cloud mass in the halo
would be  $\masscl \simeq 2 \times 10^{10} \Msun$. 
This implies $N_{\rm cl} = 4000$, consistent with the
number expected from the HIPASS count.

It  is remarkable that  the numbers in all of  these cases work out to
favor roughly the   same cloud mass,  $\mcloud \simeq  (3-5) \times 10^6
\Msun$.    This is likely  something  of a coincidence considering the
crudeness of our model.  Although we have  not focused on it here, the
HVCs are   observed   to have  a distribution    of sizes  and  column
densities.  This might be achieved by allowing a distribution of cloud
masses and some more sophisticated treatment of how they might disrupt
upon approaching  the galaxy.  Nonetheless,  we  take it  as a positive sign
that our simple model  is able to match  the rough  characteristics of
HVCs using a single cloud mass.

\section{Quasar Absorption Systems}
\label{sec:abs}

\begin{figure}
\centering 
\vspace{0pt} 
\epsfig{file=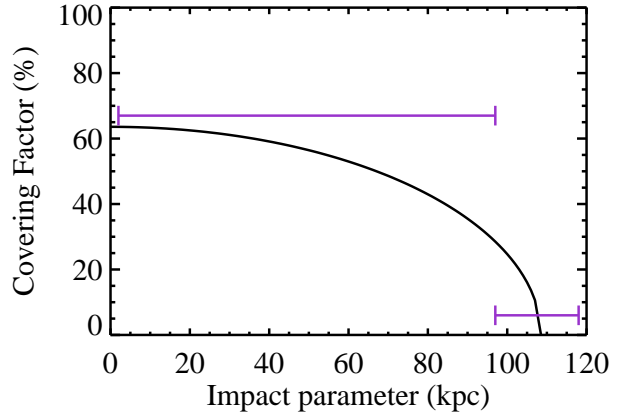,width=\linewidth} 
\vspace{0pt}
\caption{The covering factor to CIV absorbers as a function of impact 
parameter for our fiducial model galaxy at $z=0.5$ (solid line). The 
error bars are representative of the results of \citet{clw:01}.  The sharp
truncation that is indicated
by the data is roughly at $\rcool$ which is what would
be expected in our model.  More detailed comparisons are difficult because
the data spans a wide range of redshifts and luminosities.  However, with 
more data absorption systems will provide important constraints on the
properties of the warm clouds in different mass halos.
}\label{fig:abs}
\end{figure}

It has long been assumed that quasar absorption systems can
be identified with the gaseous content of galaxy halos \citep{bs:69}.  
High column density systems like Lyman limit and CIV systems 
are observed to have nearby optical counterparts for $z<1$
\citep{bb:91,clw:01,clwb:01,steid:97,lanz:95}.  Theoretically, these systems
have been modeled as arising from warm clouds embedded within a 
hot galaxy halo \citep{mm:96} in a way that is quite similar to
what we describe here.

\citet{clw:01} find in their sample that when the impact parameter of 
the quasar is $ < 97$ kpc, $67\%$ 
of galaxies show CIV absorption, while when the
impact parameter is $ > 97$ kpc, only $6\%$ of galaxies show CIV absorption
systems.  We compare this rough expectation
to the cloud covering factor as a function 
of radius calculated using our fiducial parameters for a galaxy at $z=0.5$
(Figure \ref{fig:abs}).  In our model there is also a sharp
drop in covering factor that occurs 
at 
the cooling radius of the halo.  Thus our model is in qualitative 
agreement with the observations.

To properly model quasar absorption systems one must be able to connect
galaxy luminosity and type to a halo's maximum circular velocity as a 
function of redshift.  Then each observed galaxy's gaseous halo can be
modeled and compared to observations. With a great deal more absorption
data it will be possible to constrain the masses and numbers of clouds
in a halo as a function of halo mass and redshift.  One quantity that
would be useful to know is the average relationship between the column 
density of the absorption system and the total amount of mass along 
the line of sight.  This should be possible combining weak gravitational 
lensing with a large sample of quasars and absorption systems \citep{mkbb:02},
and some progress has been made on this front \citep{mp:03}.
We discuss how other properties of absorption systems
may be useful in constraining the 
properties of the warm clouds in \S\ref{sec:imp}.

\section{The Luminosity Function}
\label{sec:lf}

One of  the fundamental  goals  in  galaxy  formation modeling is   to
understand why there  are so few galaxies  with baryonic masses larger
than $\sim 10^{11} \Msun$.  The cooling-time arguments
\citep[e.g.][]{wr:78} were   originally devised    with the  goal   of
explaining this upper-mass cutoff, but it is now accepted that the
cooling radius treatment
alone cannot   do  the  job  in  the  context  of  modern $\Lambda$CDM
cosmology \citep[e.g.][]{wf:91,tw:95, sp:99,bens:03}.  
The  $\Lambda$CDM halo  mass  function  (or  velocity
function) follows a near power-law distribution over the mass-scale 
(or velocity scale) of the Milky-Way halo
\citep[e.g.][]{gonz:00}. 
 In contrast, the luminosity
function drops quickly above the luminosity
scale of the Milky Way  \citep[e.g.][]{blan:03}.
The cooling radius treatment reduces the fraction of gas that
cools in high-mass halos, but only moderately, and  certainly not
at the level required to explain the characteristic luminosity
of galaxies (see Fig. \ref{fig:massfunc} below).

The issue is highlighted noticeably by the results of
\citet{bell:03} who used data from 2MASS and the SDSS
to construct the baryonic mass function of stars+gas in the local
universe.  They concluded that the number  density of galaxies falls off
sharply above a cold baryonic  mass of $M_*  \simeq 10^{11}
\Msun$ (shaded band in Fig. \ref{fig:massfunc}).  By integrating  
their mass function,  they found
that the total mass in cold baryons in the  local universe  is only
$\sim  10 \%$ of   the total baryonic mass   expected from BBN and the
concordance $\Lambda$CDM model.   While the original cooling arguments
suggested that most of  the baryonic mass  would  end up in  stars, it
seems now that  most of the  baryons have ended  up in hot gas, or  at
least in some state that is not associated  with central galaxies.  As
discussed by \citet{bens:03}, explaining the the sharp cutoff at the
bright end of the luminosity function is difficult within the standard
scenario without resorting to extreme conduction (above the Spitzer value) or
hot super-winds with energies beyond expectation.

\begin{figure}
\centering 
\vspace{0pt} 
\epsfig{file=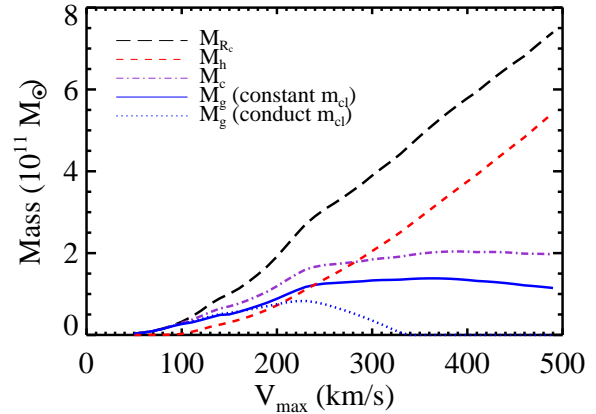,width=\linewidth} 
\vspace{0pt}
\caption{The total baryonic mass within
the cooling radius, $\massrcool$ (long-dashed line) as a function
of halo $\vmax$, assuming a gas metallicity of $Z_g = 0.1$.
Also shown is total mass that we expect to have cooled, 
$\masscool$ (dot-dashed line) 
and the total mass in the hot core (short dashed line).
The total cooled mass approaches a constant in high-$\vmax$ halos
because much of the mass within the cooling radius ends up in
a pressure-supported hot-gas core.  The
mass in cooled gas that accumulates into the
central galaxy $\massgal$ is plotted
for two different assumptions about the
cloud masses.  The solid line shows the resultant
galaxy mass calculated assuming a 
constant $\mcloud = 5 \times 10^{10} \Msun$ and the 
dotted line shows the result assuming that the cloud
mass is set by conduction (see text).
}\label{fig:uppermass}
\end{figure}

Figure \ref{fig:uppermass} shows how this problem might be alleviated
by  allowing a  two-phase medium  to  develop during the  cooling
process.  The  dashed line  shows the total  baryonic mass  within the
cooling radius, $M_{\rcool}$, as  a function of halo maximum  circular
velocity.   This mass, associated  with  the central galaxy in the
standard treatment, continues to  rise rapidly as the halo  velocity increases
and would naively lead to a population of 
giant $\sim 8 \times 10^{11} \Msun$  galaxies
associated with galaxy-group  halos   with  $\vmax \simeq  500   \kms$.  
The dot-dashed line, showing the total cooled mass,
$\masscool   = M_{\rcool} - \masshot$,   determined  by the  two-phase
treatment described in  \S \ref{sec:two-phase}, is more encouraging.
Interestingly, this mass approaches a characteristic value of
$\masscool \simeq  2  \times 10^{11}   \Msun$ for halos  with
$\vmax \gsim 250   \kms$.  

\begin{figure}
\centering
\vspace{0pt}
\epsfig{file=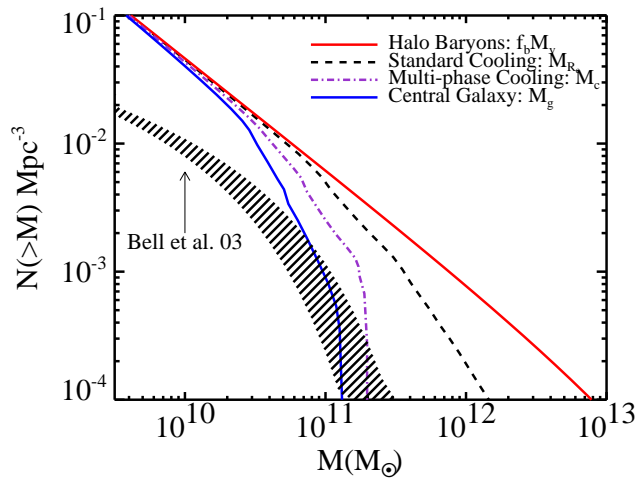,width=\linewidth}
\vspace{0pt}
\caption{The cumulative baryonic mass function of galaxies
reported by \protect{\citet{bell:03}} (shaded band)
compared to the cumulative mass function of halo baryons
(top solid line).
The short-dashed line shows the (central)
galaxy mass function that arises
from assuming all of the mass within each halo's cooling radius
cools onto the central galaxy, $\massrcool$.
The dot-dashed line is the cooled mass ($\masscool$) function that arises
in our picture, which allows for the presence of a hot
corona in each halo (see text and Fig. \protect{\ref{fig:uppermass}}).
Finally, the lowest solid line shows
the central galaxy mass function that results
from modeling the survival probability of cooled
clouds in the halo, assuming a
 typical cloud mass of $5 \times 10^6 \Msun$.
Only clouds that fall to the center of each halo
are assumed to contribute to the central galaxy.
As mentioned in the text, no merging has been accounted
for in this estimate.  Merging will tend to populate
the massive tail of the galaxy mass function, likely
bringing it even more closely in line with what
is observed.
}\label{fig:massfunc}
\end{figure}

The reason why the total cooled mass approaches a constant
in large halos is that
the  hot  halo  core,  $\masshot$,   
grows rapidly  with $\vmax$ because of the corresponding
increase in the cooling density (short dashed line).  This 
compensates for the increase in $M_{\rcool}$, resulting
in $\masscool   = M_{\rcool} - \masshot \rightarrow$ constant.
Once $\vmax \gsim 250 \kms$, $\rcool \rightarrow$constant 
(Fig. \ref{fig:rcool})
and 
$\masshot \simeq  \rho_c  \rcool^3 \propto \rho_c \propto \vmax$ 
(see equations \ref{eq:rcoolfit} and \ref{eq:rhocool} and Fig. 
\ref{fig:ncool}). 
In  this regime, the total mass inside the cooling radius for our
initial hot gas profile  also increases proportionally to $\vmax$,  so
that $M_{\rcool} \propto \vmax$  as  well.  Therefore, both the  total
mass within $\rcool$ and the mass in the hot core increase with $\vmax$ 
in the same way (the dashed lines in Fig. \ref{fig:uppermass}).  
The amount of cooled mass remains constant at the value it had when the 
slopes began to match.
Of course for different assumptions about the initial hot gas profile
the slopes may not exactly match and the amount of cooled mass may 
increase (or decrease) slightly instead of remaining constant.  However,
it still will change much less drastically than the total mass within 
the cooling radius.

The solid line in Figure  \ref{fig:uppermass} shows our expectation
for the central galaxy mass as  a function of $\vmax$.
This is determined using
the methods outlined in  \S  \ref{ssec:total}, assuming 
a constant cloud mass of
$\mcloud = 5\times 10^{6} \Msun$.  We see that in this case, the total
mass in cold gas that ends up in the central galaxy approaches a value
$\massgal \simeq 1.5\times 10^{11}  \Msun$ for halos with $\vmax \gsim
250 \kms$.  
Of course, this treatment has not included any mergers between
halos, so that this maximum galaxy mass really represents the
maximum mass galaxy sitting within a relatively
``quiescent'' halo.    An interesting implication 
is that forming a galaxy more massive than
$\sim 1.5 \times 10^{11} \Msun$ would require a merger.
This may be relevant in explaining why spheroidal galaxies tend to 
dominate the bright-end of the luminosity function.

Another possibility is that the characteristic cloud mass is
not  constant, but scales with  the halo temperature in some way. 
The short-dashed line assumes that cloud masses are set by
conduction ($\mcloud \propto T^{11/4}$) 
as described in  \S \ref{ssec:conduct} (
dot-dashed line in Fig.  \ref{fig:masses}).  
In  this case, cloud  masses
become quite large,  $\mcloud  \gsim 10^8 \Msun$  in  $\vmax \gsim 300
\kms$ halos.  The infall time  of clouds scales as $\tau_{\rm
in} \propto \mcloud^{1/3}$ (equation \ref{eq:tauram}, \ref{eq:taucc})
and thus becomes quite long at $T$ increases.  Clouds
tend  to remain in the hot gas
halo rather   than fall  in   to  contribute to  the   central galaxy
in this case.
Interestingly, massive clouds of this type will likely  be  Jeans unstable in
high-mass halos (e.g. Fig.  \ref{fig:masses}).  In this
case, they  may form stars and
become  small galaxies on their  own (see our discussion in 
\S\ref{sec:imp}). 
As in the fixed $\mcloud$
case,  giant CD  galaxies could only  form via  mergers in
this picture.

The mass function of galaxies is shown in
Figure \ref{fig:massfunc}.  The shaded band shows the baryonic
mass function of galaxies in the Universe
determined by \citet{bell:03}.  The width of the band
indicates  their uncertainty (which comes mainly from the IMF).
Compare this to the upper solid line, which
shows the halo mass function of \cite{st:99} scaled by
the mass of baryons in each halo ($\Mvir \rightarrow f_b \Mvir$).
The offset is roughly a factor of $\sim 10$ in normalization,
and from this one 
can immediately see most of the baryons in the universe
cannot be in the form of cold, galactic material.
The mass function that results from assuming that all of the baryons
within each halo's cooling radius cool onto a galaxy 
(shown by the short-dashed line) cannot solve the problem.
That is, it shows no 
sharp drop in galaxy counts  above $\sim 10^{11} \Msun$.
As expected from the above discussion, 
the ``cooled mass'' function derived using our multi-phase picture
does much better in accounting for this cutoff (dot-dashed line).  

The mass function
of gas that we expect to actually fall into the central
galaxy is shown by the lowest solid line in Figure \ref{fig:massfunc}.
In this estimate we have assumed that clouds have a typical mass of
$5 \times 10^6 \Msun$ (as in the solid line in Fig. \ref{fig:uppermass}).
The high-mass cutoff compares quite well to the data in this case.
We expect galaxies more massive than this cutoff  to be produced
solely via mergers, and we suggest that mergers will tend to populate 
the tail of the mass function above $\sim 10^{11} \Msun$, bringing it even
more in line with observations.
As is clearly seen,  our cooling scenario will not help
explain the well-known faint-end slope problem (the
low number density of
galaxies smaller than $\sim 5 \times 10^{10} \Msun$). 
Of course, feedback likely plays a major role in this regime.

It is quite clear from Fig. \ref{fig:uppermass} that the galaxy mass 
stops increasing because most of the baryons remain in the hot core.
The hot gas in higher temperature, more massive halos can be observed
with x-ray telescopes.  These observations suggest a more complicated 
picture then we have been describing here; where energy injection 
from AGN \citep[e.g.][]{omma:04} or other sources may be needed  
to explain the observed x-ray luminosity vrs. temperature relationship
\citep[e.g.][]{ms:97,wfn:00}.  Energy injection into the hot gas may also 
occur in galaxy mass halos.  We suggest that it may not be needed to 
explain the high mass cutoff in the luminosity function, although
undoubtedly heating processes occur to some extent.

\section{Implications, Observations, and Future Work}
\label{sec:imp}
Central to our model is the existence of a hot, low density medium that
surrounds galaxies --- an idea first proposed by \citet{spit:62}.  The
hot gas density we expect for a Milky-Way type system is 
$n_h = {\rho_h}/{\mup m_p} \sim 8
\times 10^{-5} {\rm cm}^{-3}$ at $\sim 100$kpc.  
Evidence that such a corona exists has been growing in 
recent years.  For example,  gas clouds in the Magellanic Stream 
are more easily understood if they are confined by a hot gas 
medium \citep{stan:02} as are the shapes of supergiant shells along the 
outer edge of the LMC \citep{deboer:98}.  
Indeed,  detection of H-alpha emission at
the leading edges of clouds in the Magellanic Stream is best explained
by ram-pressure heating  from a substantial hot gas  halo 
($n_h \sim 10^{-4} {\rm{cm}}^{-3}$) at $\sim50$kpc \citep{ww:96}. 
Unfortunately, most of the quantitative limits  on
galactic hot gas densities are  model dependent, 
but the currently limits are at least in line with our predictions:
 $n_h \lsim (1-10)\times10^{-5}$ cm$^{-3}$.
\citep[e.g.][]{snow:97,bens:00,br:00,md:94,mura:00,tripp:03}.
Interestingly, \citet{qm:01} argue that gas densities as low as
$\sim 10^{-5} {\rm cm}^{-3}$  cannot produce the head-tail  
position-velocity gradients observed by \citet{bruns:00,bruns:01} 
for $\sim 20\%$ of HVCs.  Our hot halo core is somewhat denser than this, 
so the head-tail gradients may be consistent our scenario.

One of the most promising  methods for probing the hot  gas halo is to
observe it  in absorption.   Indeed,  OVI in or around the Milky Way halo
has been detected in this manner by FUSE \citep{sava:00,semb:03a}.
Higher ionization lines (e.g. OVII and OVIII), 
can be observed by XMM-Newton and Chandra.
\citet{nica:02} reported the first detection of
highly ionized  oxygen and neon in proximity to the Local Group 
and similar detections have followed. 

There is some debate over exactly where this absorption takes
place and how the observed OVI and OVII absorbers are related
\citep[e.g. compare][]{semb:03b,nica:03b}.
In the context of the corona we predict,
measurements to known
sources in the Local Group may help avoid confusion.
For example, our fiducial model predicts a
column density in Hydrogen of $N_H =10^{19} \cm2$ 
along a line of sight to the LMC, assuming a distance of
50kpc.  With $Z_g=0.1$,  this gives a total expected column 
density in oxygen of $5.1 \times 10^{14} \cm2$.
At the temperature and density
we expect for the Milky Way corona, most of the oxygen should
be in the form of OVII \citep[e.g.][]{mwc:03},
but observing multiple elements and ionization
lines would be useful for constraining
the precise ionization state, temperature,
and density of the hot medium(s) responsible for any absorption.

A related test  will come from searches for 
metal lines associated with  HVCs.  \citet{semb:03a}
have argued that high-velocity OVI features observed by FUSE highlight
the boundaries between warm clouds of gas and  a highly extended, hot,
low-density corona around the Galaxy.  A related analysis suggests
that these  systems  are associated  with   HVCs \citep{tripp:03}.  We
would expect just this situation in our model.  Note that
\citet{nica:03a} have argued that the high-velocity
OVI absorbers are better-described by a Local Group population,
but they also allow for the possibility that they trace
an extended Galactic corona.  This second interpretation is in line
with our expectations.

The soft x-ray  background provides a  less-direct  method for probing
the  hot gas cores  of galaxy halos.  In our  model, the hot-gas cores
are expected to be of  low  density, and to   have a rather low  x-ray
surface  brightness, not    directly  detectable  by current     x-ray
satellites.  However, this hot  gas will contribute to  the soft-x-ray
background.  Predictions for the  contribution to the soft-x-ray
background may provide interesting limits on the model.

More detailed comparisons with existing and future HVC data will require
more realistic models of clouds in a hot halo.
The  properties of clouds will need to  be  modeled in the
presence of an ionizing background field and interactions with the hot
gas background should  be properly taken  into account.   For example,
\citet{wvw:02}
have  argued that  measurements of  H-alpha emission in  HVCs can  put
constraints on their distances as long as H-alpha recombination is
caused  by  photoionizing  radiation  from the   Milky  Way.   Another
possibility  is that the H-alpha recombination is due to
collisional ionization caused by ram pressure interactions with the hot 
gas halo.  Such a scenario would be consistent
with the OVI observations discussed above.  Further, the ambient
pressure from the hot gas halo is expected to vary as a function of
radius from the halo center,
and  this  would lead to  varying cloud
sizes (and  densities) at fixed mass.   The spectrum of  HVC sizes and
column densities might then be used to constrain the nature of the hot
gas corona, and even  to test the mass spectrum  of clouds.  Of  course,
without knowing  the distances  to individual HVCs,  this can  only be
done in a statistical sense.

Searches for clouds around other galaxies will help establish whether
clouds of the type we discuss are as common and numerous as we expect,
and might even be used to test how 
cloud masses and hot core properties
vary with halo mass and galaxy luminosity.
\citet{pisano:04} recently performed a search for HI clouds around 
three nearby galaxy groups and
found that if a population of HI clouds exists around galaxies
like the Milky Way, they must be clustered within
160 kpc and have HI masses  $\lsim 4 \times 10^{5} \Msun$.
The clouds we expect are consistent with these limits, but
should be detectable if the detection limits are relaxed only
slightly.
\citet{thil:04} have discovered a population of $\sim 50$ HI clouds around 
M31, with HI masses of $\sim 10^6 \Msun$.  Our model would
suggest that these are likely the most massive of the many thousands of 
clouds that should surround M31. 
We predict that deeper surveys with better angular resolution will
find these clouds. 

Quasar absorption systems provide another important avenue 
for determining the properties of warm clouds.  While cloud
populations of this type cannot be studied in detail,
the study of absorption systems can probe
wide range of halo types.  A
cross-correlation between absorbers and galaxies may
yield useful information on cloud sizes, densities, and covering 
factors.
In the future, a cross-correlation survey, similar to 
that of \citet{clw:01}, 
but utilizing a large optical survey, would
yield tight constraints on the distribution and scaling of 
cloud masses.  

Another advantage of absorption systems is that they probe the gaseous
halos of  galaxies   at early  stages of formation,
$z  \sim  3$.   Multi-phase  cooling  may  be a  crucial
ingredient   in  understanding   the properties    of  these systems.
\citet{mpsp:03} pointed out that a large fraction of the halo gas must
be in form of warm clouds in  order to explain the observed kinematics
of the high-ion component in damped Lyman alpha systems \citep{wp:00a}.
Including  the multi-phase medium  self-consistently will be important
for   precise  comparisons  with this  data.    Indeed, damped systems
themselves show complex kinematics  \citep{pw:97,pw:98} that cannot be
explained in CDM cosmologies without the presence of a large amount of
gaseous substructure \citep{hsr:98,mm:99,mpsp:01}.  Warm clouds may
make an important contribution to this substructure, possibly after
they are disrupted by tidal forces.

If pressure-supported clouds exist in the halos of most galaxies, this
could be important for interpreting the 
flux ratios of multiply-imaged quasars \citep{mm:01,dk:02,mm:03}.   
The flux ratio anomalies have been used
to argue the existence of the low-mass dark matter halos
predicted by $\Lambda$CDM N-body simulations
\citep{moore:99a,klypin:99b}.  However, there are some indications
that the fraction of mass  in low-mass $\sim 10^7 \Msun$ substructures
is even higher than predicted \citep{zb:03,mm:03}.
The warm clouds we have described here will also cause
fluctuations in the gravitational potential, and  may  be important.
Indeed, the  mass  fraction in  clouds  is expected  to  be at the few
percent level, and   this is  roughly the   same  as the dark   matter
substructure population.   Fortunately, the  warm clouds may   also be
detected by absorption in the quasar spectra so  it may be possible to
disentangle the two signals.  Also,  since the clouds are expected to
be roughly constant density,  they may not provide as strong a signal
as the more concentrated dark matter clumps.

Multi-phase cooling might also help resolve the long-standing
problem of forming disk galaxies without angular momentum loss
in cosmological simulations \citep[e.g][]{ns:00}.
Specifically, if cooled gas remains in warm clouds instead of settling
into the  galaxy,  then those  clouds  can  retain  and  gain  angular
momentum during  mergers.  When the  clouds  eventually fall  in, they
will produce large disks \citep{md:02}.  This scenario    would be
especially   helpful if angular   momentum in  dark   matter  halos is
predominately acquired in mergers \citep{mds:02,vitv:02}.
Interestingly, \citet{rysh:04} showed that by allowing a cold/warm
medium to exist within  cooled, star-forming material,  they could
improve the likelihood of disk formation in cosmological simulations.
Specifically, the disk is more stable to its own self gravity, and less
likely to fragment and loose angular momentum after it forms.
Of course, this effect will only help if the material that forms 
the disk initially retains a large amount of angular momentum.
It is in the retention of halo angular momentum that the warm/hot
cloud picture becomes important.
Therefore,  a full multi-phase approach, with an allowance
of both  cold/warm and warm/hot phases, could
 lead to more success in this direction. However, as
we mention in \S\ref{sec:conc}, there are significant
computational challenges to overcome.

Finally,  as
seen in Fig. \ref{fig:masses}, the Jeans mass for a cloud decreases as
a function of  halo temperature, making it more likely that a cloud
will collapse under its own gravity in high-temperature halos.
This possibility will be more likely if
the typical cloud mass increases in high-temperature halos.
This is  what  is expected, for example, if conduction sets the
characteristic cloud mass as discussed in \S \ref{ssec:conduct}.  If a
cloud's mass exceeds the Jeans mass then it will
likely fragment  to form stars.   It   is perhaps to be
expected then that there should exist 
a  population of low-mass galaxies, born of
fragmented gas in clusters, with  no associated dark matter. 
Dwarf galaxies of this type  would likely be younger than
other low-mass spheroids, and have relatively low mass to light ratios
by comparison.

\section{Conclusions}
\label{sec:conc}

In  this paper  we have  taken  a step  towards   modeling the complex
realities of   astrophysical  hydrodynamics using a    simple analytic
treatment that allows  the development of  a two-phase warm/hot medium
during gas cooling.   Appealing to standard cooling instability
arguments, we showed that  if  cooling proceeds  by the
formation of   warm clouds   embedded within   a  low-density  hot gas
background   then this can
explain the 
characteristic upper limit  in the observed 
baryonic masses of galaxies,   $\sim 10^{11}
\Msun$ (\S \ref{sec:lf}).  
In the standard treatment,  all of the
mass within the cooling radius of each halo cools onto the 
central galaxy, while our approach 
allows the survival of a hot
gas core with a density close to the cooling density 
in each the halo.  The fraction of mass that remains
in the hot core component is large in
high-mass halos because the cooling density is high,
and this gives rise to an upper-mass limit
in cooled material in these systems.

When 
applied to Milky-Way size halos, the standard single-phase treatment
over-predicts the Milky Way  mass by more  than a factor of two, while
our multi-phase treatment helps explain  the   Milky Way galaxy  mass
naturally,   without     the  need   for  excessive      feedback  (\S
\ref{sec:infall}).  Because of the thermal instability, 
we argue that galaxy formation should proceed via the infall of warm,
pressure-supported clouds.   Further,   if the  typical cloud mass   is
$\mcloud \simeq 5 \times  10^{6} \Msun$, the residual cloud population
is significant,  and  we identify  these pressure-supported  fragments
with the observed High Velocity Cloud population of  the Milky Way (\S
\ref{sec:hvc}).   The typical Galacto-centric  distance to HVCs in our
picture is set  by the cooling  radius, $\sim 100$kpc.  The same cloud
mass helps explain the baryonic mass of the Milky Way, can account for
high-ion  absorption systems  in  distant galaxies (\S \ref{sec:abs}),
and allows clouds to survive destructive processes in the halo (\S
\ref{sec:cloud-masses}).

Including the multi-phase treatment presented here in
standard semi-analytic models
should be straightforward.  Currently this type of
modeling tracks two phases of gas: the mass contained in
hot halo gas, and the mass in the ``cold'' central galaxy.
To include the
multi-phase  medium, one must simply add the hot core from
equation~\ref{eq:mcool} to the gas in the hot phase, and include  an
additional    accounting   for    warm  cloud material.     As sketched   in
equation   \ref{eq:massbudget}, gas first  cools into  warm clouds and warm
clouds become deposited in the central galaxy on an infall timescale.
Once clouds fall in, the typical recipes for star formation may
be applied, although it is likely that less feedback will be needed.

Comparison with cosmological hydrodynamic simulations may be
somewhat more difficult.  At present, these simulations do not resolve
the multi-phase structure  of  halo gas,  and this  may lead  them  to
predict infall rates of cooled gas that are
similar to  those expected  from simple
cooling radius arguments.  In order to  test these expectations,
simulations would need to resolve typical cloud masses of $\sim 10^6
\Msun$ as well as properly follow cloud fragmentation in the diffuse
gas halo.  \citet{yepes:97} and later
 \citet{sh:03b} have in fact included a sub-grid model
for multi-phase gas in their SPH codes, but this applies only
to the star-forming, cold/warm medium.  This approach
seems to alleviate many of the problems faced by similar
codes in the past \citep[e.g.][]{rysh:04}, but not  the 
``over-cooling'' problem discussed here.  
If the computational challenges can be overcome, hydrodynamic simulations 
with full multi-phase cooling may yield even more 
encouraging results.

In conclusion, multi-phase cooling is expected on theoretical grounds
and can alleviate many of the problems that 
arise in the standard, single-phase procedure.  The fact that warm clouds
of the type predicted seem to be observed
only further enforces the relevance of adopting this approach
in models of galaxy formation.
If we are correct, then
the High-Velocity Clouds of the Milky Way
are tracers of the fundamental fuel supply that governs
galaxy formation in the Universe.  Set in this context, the study of 
hot gas, HVCs, 
and their counterparts in external galaxies will have 
significant impact on how we understand galaxies and their assembly.

\section*{Acknowledgments}
This work has benefited from useful conversations with
A. Babul, L. Blitz, N. Katz, A. Kravtsov, S. Mather, B. Robertson, 
J. Simon, 
T. Tripp, R. Wechsler and B. Weiner. We thank J. Miralda-Escud{\'{e}} for
for encouragement and for pointing out an error in Section 5 of an
earlier draft.  D. Weinberg provided sage advice on the introduction
and several useful suggestions.  We thank the anonymous referee for 
comments leading to an improved paper. 
 We gratefully acknowledge the
staff and general management of the Sababa Hotel in the Sinai, Nuweiba for
its generous hospitality during the conception and early
stages of this work.  AHM is supported by NSF grants AST-0205969
and  AST-9802568. JSB is supported by NASA through  Hubble
Fellowship grant    HF-01146.01-A  from the  Space   Telescope Science
Institute, which  is operated by  the Association  of Universities for
Research in Astronomy, Incorporated, under NASA contract NAS5-26555.
\bibliographystyle{mn2e}         
 
\bibliography{me,gf,abs,cosmo,reion,dm,hydro,spin,clouds,cond,gals,xray,lens}

\appendix

\section{A. The cooling function}
\label{sec:coolfunc}

\begin{table}
\begin{center}
\begin{tabular}{ccccc}
\hline
$Z_g$  & $T_b (10^6 K)$ & $\vmax(T_b)$ & $\alpha$ & $\Lzz$\\
\hline
\hline
0.0   &  1.0    &  205  & -0.80   &  0.19 \\
0.03  &  2.4    &  307  & -0.17   &  0.45 \\
0.1   &  4.2    &  394  &  0.23   &  1.0  \\
0.3   &  7.7    &  518  &  0.60   &  2.0  \\
1.0   &  16.    &  722  &  1.00   &  4.4  \\
\hline
\end{tabular}
\caption{Properties of the cooling function for several
different choices of hot gas metallicity, $Z_g$, 
as parameterized by the fitting function
in equation \ref{eq:coolfunc}.  The temperature $T_b$ 
is the temperature above which Bremsstrahlung radiation begins
to dominate cooling, and $\vmax(T_b)$ is the corresponding
halo maximum circular velocity.  The parameter $\alpha$ represents
the fitted slope to the cooling function in the regime:
$T_{r} < T < T_{m}$, where $T_r = 1.5 \times 10^4$K and
$T_m \simeq 1.5 \times 10^5$K
for $Z_g > 0$ and formally $T_m = T_b$ for $Z_g = 0$. 
Finally, $\Lzz$ is the value of the cooling function in units
of $2.6\times 10^{-23} {\rm cm}^{3} {\rm erg} \,{\rm s}^{-1}$
as defined in equation \ref{eq:Lfit}.
}\label{tab:cf}
\end{center}
\end{table}

Astrophysical plasmas with temperatures greater than $10^4 K$ primarily 
cool by radiative processes.  The cooling function $\Lambda(T)$ can be
calculated \citep[e.g.][]{sd:93} as a function of the gas metallicity $Z_g$.  
It is useful to introduce the dimensionless cooling function $\L23(T)$:
\beq
\Lambda(T) \equiv 10^{-23} {\rm cm}^{3} {\rm erg} \,{\rm s}^{-1} \L23(T)
\eeq 
Figure \ref{fig:coolfunc} shows
$\Lambda(T,Z_g)$ plotted as a function of temperature for five example
metallicities.  Note that for galaxy-sized halos ($\vmax\simeq 100-200 \kms$)
and mildly-enriched gas, the dimensionless cooling function takes values of 
order unity.

We can approximate the cooling function 
by a series of power-law fitting functions that captures the 
important scalings.  For zero metallicity gas there are two
important temperature regimes: $T >  T_{b}$, when 
the dominant cooling process is Bremsstrahlung radiation, 
and $T_b > T > T_{r}=1.5 \times 10^4$K,
where the dominant cooling process is the recombination of hydrogen.
For enriched gas, there is a third important temperature
scale, $T_m$, where metal line cooling becomes 
important.  The Bremsstrahlung region becomes important at
higher temperatures for more metal rich gas, and we find that
to good approximation $T_{b}=10^6+1.5 \times 10^7 {Z_{g}}^{2/3}$K.
We also adopt $T_m = 1.5 \times 10^5$K.

The cooling curve is then given by,
\begin{eqnarray}
\label{eq:coolfunc}
\L23^f(T) &=& \quad 12 \quad \lp{{T}\over{T_{r}}}\rp^{\alpha}  \hspace{1cm}  
T_{r} < T \le T_{m}
\nonumber \\
\L23^f(T) &=& \L23^f(T_{m}) \lp{{T}\over{T_{b}}}\rp^{-1} \hspace{1cm}   
T_{m} <  T  \le T_{b}
\nonumber \\
\L23^f(T) &=& \L23^f(T_{b}) \lp{{T}\over{T_{b}}}\rp^{1/3} \hspace{1cm}  
T > T_{b},
\end{eqnarray}
where $\alpha=1-\frac{1}{3}\ln{Z_g}$ ($Z_g \ne 0$),  and we have added
the superscript $f$  to indicate that this is a fit to the true cooling
function.  In  the case of zero  metallicity gas $\alpha=-0.8$  and the
middle expression is not used  (that  is  $T_{m}  =  T_{b}$). Values
of $\alpha$ and $T_b$ for some example metallicities can be found 
in Table \ref{tab:cf}.  This
approximation is at worst good to within a factor of $3$ and is much
better than that for most metallicities and temperatures.

\section{B. Hydrostatic Equilibrium Solutions}
\label{sec:hydrostatic}

Here we derive the hydrostatic equilibrium solutions for 
gas profiles, assuming that the gravitational potential of the
system is dominated by an NFW background halo.
The hydrostatic force balance equation is
\beq
\frac{dP}{dR} = \frac{-V_c^2(R) \rho(R)}{R},
\eeq
where $R$ is the distance from the center of the spherically-symmetric
halo.
The density follows that
given by equation \ref{eq:nfw}, and the implied rotation curve follows
\beq
V_c^2(R) =  \frac{G M(R)}{R} = 9.26 c_g^2 \frac{R_s f(R/R_s)}{R},
\eeq
where $f(x) \equiv \ln(1+x) - x/(1+x)$, and we adopt
$c_g = \vmax/\sqrt{2}$ as the sound speed written
in terms of the maximum circular velocity.

If we assume $P = K_\gamma \rho^\gamma$
then $K_\gamma = c_g^2 \rho_c^{1-\gamma}$ and
the solution to the hydrostatic equation for
$\gamma \ne 1$ is
\beq
\rho_g(x) = \rho_c \lb 1 +  \frac{9.26(\gamma - 1)}{\gamma}
\lp  \frac{\ln(1+x)}{x} -  \frac{\ln(1+\ccool)}{\ccool} \rp \rb^{\frac{1}{\gamma -1}}.
\eeq
For $\gamma = 1$ the solution is
\beq
\rho_g(x) = \rho_c \exp \lb \frac{9.26}{x}\ln(1+x) - \frac{9.26}{\ccool}\ln(1+\ccool)\rb,
\eeq
where $\ccool \equiv \rcool/R_s$.  
In writing this solution, we have required that the hot gas density
at the cooling radius, $\rcool$, is equal to the cooling density, $\rho_c$.
For the adiabatic assumption adopted in the paper, $\gamma = 5/3$
and the solution for pressure, density, and temperature becomes
\beqa
P_g(x) & = & P_c \lb 1 + \frac{3.7}{x}\ln(1+x) - \frac{3.7}{\ccool}\ln(1+\ccool) \rb^{5/2}, \\ \nonumber
\rho_g(x) & = & \rho_c \lb 1 + \frac{3.7}{x}\ln(1+x) - \frac{3.7}{\ccool}\ln(1+\ccool) \rb^{3/2},  \\ \nonumber
T_g(x) & = & T \lb 1 + \frac{3.7}{x}\ln(1+x) - \frac{3.7}{\ccool}\ln(1+\ccool) \rb .
\eeqa

For our adopted adiabatic profile, 
the pressure,  temperature, and  density increase slowly  towards
the center of the halo, reaching  core values
as $x \rightarrow 0$: $T_0 = \alpha T$, $\rho_0 = \alpha^{3/2} \rho_c$,
and $P_0 = \alpha^{5/2} P_c$ with
$\alpha = 4.7 - 3.7  \ccool^{-1} \ln(1+\ccool)$.  For the typical
range $\ccool = 2-20$ we find a rather modest range of values
$\alpha \simeq 2.7 - 4.1$. 
It is worth pointing out
that while the central 
density is higher than what we have defined as the
``cooling density'' (set at $T=T_c$), it does not imply
that the system is drastically unstable to cooling.
Indeed, the central hot gas is likely to have reached its
state of density and pressure more recently than the non-radiating
gas at large radii.  Moreover,
for $T_m < T < T_b$, the local cooling time will scale as 
$\tau_{\rm cool} \propto T/\rho \Lambda(T) \propto T^2/\rho \propto \alpha^{1/2}$, and thus the central gas will cool  more slowly than the
outer halo gas.

Most of the gas, by volume, 
is quite close to the state at $R_c$.
The total mass in the form of hot gas takes the form
\beq
M_{\rm h} = \frac{4 \pi}{3} r_c^3 \rho_c \etarho(\ccool)
\eeq
where the function $\etarho$ is
determined by numerical integration to be well fitted
by
\beq
\etarho(c) \simeq 1.42c^{0.3} \lb 1 + (c/1.65)^{1.7}\rb^{-0.24},
\eeq
which is good to $<1\%$ for $c=1-20$, and is a rather weak function of
$\ccool$: $\etarho  \simeq  1.45-1.25$ for $\ccool =   1- 20$.  Clearly  the
average  gas density  within  $r_c$ will   be  $\bar{\rho_h} = \etarho
\rho_c$.   In   the main part   of   the  paper   we  work under   the
approximation that  $\etarho$  is a constant  and adopt  a typical value of
$\etarho = 1.35$.

Similarly, we can estimate the volume-averaged pressure of the gas
within $\rcool$:
\beq
\bar{P_{\rm h}} = \frac{4 \pi \int_0^{\rcool} P(r) R^2 dR}{4 \pi \rcool^3/3}
= P_c \etap(\ccool),
\eeq
where the function $\etap$ is
found to be well-fitted by
\beq
\etap \simeq  2.5c^{0.33} \lb 1 + (c/3.9)^{2.9}\rb^{-0.24},
\eeq
and spans the range $\etap \simeq 2.1-3.3$ (for $\ccool = 1- 20$). 
As with the gas density, in the main part of the paper we work under
the approximation that the hot gas pressure can be well-represented
by a constant, with $\bar{P_{\rm h}} = \etap P_c$, and adopt the
typical value $\etap \simeq 2.7$.   The temperature profile
is relatively flat, and we adopt $\bar{T_{\rm h}} = T$ throughout the
paper.

\end{document}